\documentclass[prb,preprint,showpacs]{revtex4}
\usepackage{epsfig}

\begin{document}

\draft

\title{Structural and energetic properties of nickel clusters: 
$2 \le N \le 150$}

\author{Valeri G.\ Grigoryan\footnote{Corresponding author. e-mail:
vg.grigoryan@mx.uni-saarland.de} and Michael Springborg\footnote{e-mail:
m.springborg@mx.uni-saarland.de}}
\affiliation{Physical Chemistry, University of Saarland,
66123 Saarbr\"ucken, Germany}

\date{\today}

\begin{abstract}

The four most stable structures of Ni$_N$ clusters with $N$ from 2 to 150 have 
been determined using a combination of the embedded-atom method in the version 
of Daw, Baskes and Foiles, the {\it variable metric/quasi-Newton} method, 
and our own {\it Aufbau/Abbau} method. A systematic study of energetics, 
structure, growth, and stability of also larger clusters has been carried
through without more or less severe assumptions on the initial geometries
in the structure optimization, on the symmetry, or on bond lengths. 
It is shown that cluster growth is predominantly icosahedral with 
$islands$ of {\it fcc}, {\it tetrahedral} and {\it decahedral} growth. 
For the first time in unbiased computations it is found that  
Ni$_{147}$ is the multilayer (third Mackay) icosahedron. Further, we 
point to an enhanced ability of {\it fcc} clusters to compete with 
the icosahedral and decahedral structures in the vicinity of $N=79$.
In addition, it is shown that conversion from the {\it hcp}/anti-Mackay
kind of icosahedral growth to the {\it fcc}/Mackay one occurs within a 
transition layer including several cluster sizes. Moreover, we present and apply 
different analytical tools in studying structural and energetic properties of
such a large class of clusters. These include means for identifying the overall
shape, the occurrence of atomic shells, 
the similarity of the clusters with, e.g., fragments of the {\it fcc}
crystal or of a large icosahedral cluster, and a way of analysing whether the 
$N$-atom cluster can be considered constructed from the $(N-1)$-atom one by
adding an extra atom. In addition, we compare in detail with results from 
chemical-probe experiment. Maybe the most central result is that first for 
clusters with $N$ above 80 general trends can be identified.

\end{abstract}


\maketitle

\section{Introduction}

Clusters are important materials both from the point of view of basic 
research and from an application 
point of view. Their, partly controllable, unique physical and chemical 
properties can be related to the large 
surface-to-volume ratio or to finite-size or quantum-confinement effects.
These properties make them interesting for 
use in, e.g., nanoelectronics and in catalysis.

The central issue is to understand and predict how the properties of interest depend on the 
size of the cluster. However, addressing this question is complicated by a number of
serious problems. First, clusters contain typically from some 10s to several 100s of atoms,
so that they are neither small, well-characterized molecules, nor macroscopic, approximately
infinite crystals. Moreover, in experimental studies, one often considers an ensemble of more
or less mono-disperse clusters whose size distribution is largely unknown. Second, experimental
and theoretical studies focus often on different systems: in experiment the clusters may be 
dispersed in some solvent and may possess surfactants and only sometimes are produced 
in the gas phase, whereas isolated clusters 
in this gas phase almost exclusively are the subject of theoretical studies. Third, in theoretical
works the fairly large size of the individual clusters makes it necessary to invoke one or
more approximations: either highly accurate methods on selected clusters with pre-chosen
structures are applied, or less accurate, parameterized methods are used on more different
clusters, but also then the structures are often chosen `reasonably'.

In this work we shall present results of our theoretical studies on Ni$_N$ clusters with $N$ 
up to 150. Our approach is based on the approximate embedded-atom method (EAM) in calculating 
the total energy of a given structure and we 
shall show that this method provides a good compromise between
accuracy and computational speed. Moreover, we have used our own {\sl Aufbau/Abbau} method
in determining the structures of the lowest total energies. Ni clusters provide one of the
most studies sets of clusters and, therefore, is an excellent system for exploring new
theoretical or experimental approaches. This is our main reason for focusing on that system.
Moreover, through our unbiased structure-optimization approach we are able to identify
the structures of more low-energy structures for each size, and, simultaneously, 
obtain structure and total energy for all values of $N$ up to $N=150$, whereby 
size-specific properties can be extracted.

Once the structure of a given cluster has been chosen, the variational principle allows for
a systematic improvement of the calculated electronic properties. A similar approach does not
exist for the structural properties where one, instead, has to use more or less biased
methods for determining the structure of the lowest total energy. Here, three issues make
such theoretical studies very demanding: first, the computational efforts for a single
structure scales with $N$ to some power from 2 and upwards; second, the number of structural
parameters that have to be determined scale linearly with $N$; and, third, the number of 
meta-stable structures scales faster than any power of $N$.\cite{tsai93} 
Therefore, any theoretical study
employs one or more approximation that may or may not be crucial. 

Accordingly, parameter-free calculations on Ni$_N$ clusters\cite{ca99,mi01,re95,na96,de98,kr00}
have been applied mainly for smaller clusters as well as on single, high-symmetric, larger
ones. Alternatively, unbiased structure optimizations for larger clusters are all based
on more approximate total-energy methods like the EAM or related
methods,\cite{vl92,re93,mo94,mo96,ga00,gu92,je93,bo01,st92,we96} whereby only for $N\le55$
the structure has been optimized completely unbiased, or the very simple $n$-body 
Gupta,\cite{je91,ga92,lo94,ag98,mi99} Sutton-Chen,\cite{do98} Finnis-Sinclair,\cite{na97}
Murrell-Mottram,\cite{ll00} or Morse potential\cite{hu96} has been used, whereby unbiased
structure optimizations up to $N=80$ have been carried through.\cite{do98} Only for
very simple and material-unspecific potentials, like the Lennard-Jones potential, largely
unbiased structure optimizations up to $N\sim 150$ have been carried through.\cite{northby}
Finally, Ni$_N$ clusters
have also been studied with a tight-binding method,\cite{la96} whereby the structure only 
for clusters with $N\le10$ was optimized unbiased. 

Of the experimental studies on Ni$_N$ 
clusters,\cite{pa91,pa94,pa95,pa95a,pa97,pa98,pa01,pe94,kn01} the chemical-probe
experiments\cite{pa91,pa94,pa95,pa95a,pa97,pa98,pa01,kn01} 
have given very valuable information
on the structure of the smaller clusters. 
On the other hand, results for medium and large Ni clusters,\cite{pe94}
obtained by performing near-threshold photoionization and time-of-flight 
mass spectroscopy can be used in identifying particularly stable clusters and,
subsequently, in providing information on growth modes. 

>From the earlier studies on nickel clusters it has been found 
that highly stable clusters occur for $N=13$ and $N=55$ for which the structures 
are multilayer icosahedra, i.e., the first and second Mackay icosahedra. Moreover, in several 
theoretical studies\cite{we96,mi99,do98} a {\it fcc} truncated octahedron was found 
for Ni$_{38}$ which is in accord with experimental results. Thus, it can be assumed
that for these three cluster sizes a `structural consensus' has been achieved, which
in turn provides a useful check for any further study. On the other hand, for $N=75$
some theoretical studies\cite{do98,mi99} have predicted Marks decahedron to be the
most stable structure, in disagreement with experiment. Moreover, 
the cluster growth pattern for intermediate sizes, i.e., for 
$13 < N < 55$ and for $N>55$, remains unclear and the obtained results seem to
depend strongly on the used approach. 

In the present study the structure and the energetics of the four most stable 
isomers for small and medium-sized Ni$_N$ clusters with $N$ from 2 to 150 have 
been determined for each cluster size using a combination of the EAM (for calculation of the 
total energy for a given structure), the 
{\it variable metric/quasi-Newton} method (for the determination of local total-energy
minima), and our own {\it Aufbau/Abbau} method (for the determination of the 
global total-energy minimum). Except for two cases ($N=75$ and $N=104$; here we included 
the decahedron as a starting configuration) our structure-determination
is completely unbiased. In particular, as we shall see, it is important to go beyond
$N=80$ (the upper limit in the earlier unbiased structure-optimizations) in order to
identify growth patterns. In addition, we shall present results for clusters that are
obtained as relaxed spherical parts of the {\it fcc} crystal with $N$ up to some 1000s. 

The paper is organized as follows. In Sec.\ II we briefly outline the embedded-atom method,
and in Sec.\ III we present our structural-determination methods. The main results are
given in Sec.\ IV, and a brief summary is offered in Sec.\ V. Finally, for the sake of
completeness we mention that some preliminary results on clusters with up to $N=100$ and 
results for the energetically lowest isomer were presented earlier\cite{pccp01,cpl03}
and in the discussion of the results we shall use eV and \AA\, as units of energy and
length, respectively.

\section{Embedded-atom method}

We use the EAM of Daw, Baskes, and Foiles \cite{da83,da84,fo86,da93}
for the calculation of the total energy of a given cluster with a given 
structure. The main idea of Daw and 
Baskes\cite{da83,da84} is to split the total energy of a (metallic) system of interest into
two components. The embedding energy is obtained by considering 
each atom as an impurity embedded into a host provided by the rest 
of the atoms. The remaining part is written as a sum of pair potentials. Accordingly,
\begin{equation}
E_{\rm tot}=\sum_i F_i(\rho_i^h) + {1 \over 2} \sum_{i,j \,(i\ne j)}\phi_{ij}
(R_{ij}),
\label{1}
\end{equation}
where $\rho_i^h$ is the local electron density at site $i$,
$F_i$ is the embedding energy, i.e., the energy required to embed an atom
into this density, and $\phi_{ij}$ is a short-range potential between atoms $i$ and $j$ 
separated by the distance $R_{ij}$,
\begin{equation}
\phi_{ij}={Z_i(R_{ij})\,Z_j(R_{ij}) \over R_{ij}}.
\label{2}
\end{equation}
Here the effective charges $Z_{i}(R_{ij})$ and $Z_{j}(R_{ij})$ 
depend on $R_{ij}$. 
The local electron density at site $i$ is assumed being a superposition 
of atomic electron densities
\begin{equation}
\rho_i^h=\sum_{j \,(\ne i)} \rho_i^a (R_{ij}),
\label{3}
\end{equation}
where $\rho_i^a (R_{ij})$ is the spherically averaged atomic electron 
density provided by atom 
$j$ at the distance $R_{ij}$. 

In the present approach, the atomic densities have been taken 
from Hartree-Fock calculations of Refs.\ [\onlinecite{cl74,mc81}]. Moreover, 
adjustable parameters that define 
$F_i$ and $\phi_{ij}$ have been obtained by fitting to known bulk 
properties such as sublimation energy, lattice constant, the heat of
solution of binary alloys and, additionally, to the universal equation 
of Rose,\cite{ro84} which describes the sublimation energy of the most metals 
as a function of lattice constant. The values  for $\rho_i^a$, $F_i$ and 
$Z_i$ that we have used are available in numerical form for Ni, Cu, Ag, Au, Pd and 
Pt.\cite{internet} 

Our main reason for choosing the EAM is that the EAM provides a
computationally efficient, parameterized many-particle method that allows for many 
calculations of also larger systems. Thus, with this method we can perform  
unbiased structure optimizations for clusters with well above 100 atoms, which is 
not possible using more accurate, parameter-free methods. Moreover, by comparing
(see Sec.\ IV A) with results of more accurate studies for the smallest clusters we can 
estimate the accuracy of the method. Since the method has been developed first of
all for macroscopic metallic systems, in particular the smallest clusters may be those
for which the largest inaccuracies show up.

\section{Optimization of the structure} 

Using expression (\ref{1}) we can calculate the total energy of
any cluster with any structure as a function of atomic coordinates
$\{ {\vec R}_i\}$, $E_{\rm tot}({\vec R}_1,{\vec R}_2,\, \dots \, ,{\vec R}_N)$. 
In order to obtain the closest local total-energy minimum we use the 
{\it variable metric/quasi-Newton} method.\cite{numer92} We found that this
was significantly more efficient than 
the {\it conjugate gradient} method.

For searching the global minima we have developed our own {\it Aufbau/Abbau}
method. It consists of the following steps:

1) We consider two cluster sizes with $N$ and $N+K$ atoms with $K\simeq 5-10$. 
For each of those we study a set of randomly generated structures, $N_{\rm ran} 
\simeq 1000$. Using the {\it quasi-Newton} method the $N_{\rm ran}$ relaxed structures
are identified and the structures of the lowest total energy selected. 
Each of the $N_{\rm ran}$ starting structures for a cluster with $M$ atoms 
was generated using a
random-number generator for positions within a sphere or a cube of volume $V_{cl}=
(p\cdot b_{nn})^3M$, where $b_{nn}=2.49$ {\AA} is the nearest-neighbor 
distance of bulk Ni and $p=0.8,\:1.0,\:1.2$, i.e. we considered slightly 
compressed, normal, and slightly expanded structures. We included the constraints
that the smallest allowed inter-atomic 
distance was 0.5$b_{nn}$ and each atom has to interact with at least two others. 

2) One by one, each of the $M$ atoms is displaced randomly, and the closest local 
minima is determined. If the new structure has a lower total energy than the original
one, this new one is kept, and the old one discarded. This is repeated approximately 
$500-1000$ times (depending
on cluster size). 

3) This leaves us with two `source' clusters, Ni$_N$ and 
Ni$_{N+K}$ with their lowest total energies. One by one an atom is added 
at a random position to the structure with $N$ atoms 
(many hundred times for each size), and the structures are relaxed. In parallel, 
one by one an atom is removed from the structure with $N+K$ atoms --- for each 
intermediate cluster with $N'$ atoms we consider {\it all} $N'+1$ possible 
configurations, that one can obtain by removing one atom from the Ni$_{N'+1}$ cluster. 
>From the two series of structures for $N\le M\le N+K$ those structures of the 
lowest energies are chosen and these are used as seeds for a new set of 
calculations. First, when no lower total energies are found, it is assumed 
that the structures of the global-total-energy minima 
have been identified, and we proceed to larger clusters. 

Moreover, by keeping information on not only the single energetically lowest isomer,
but more low-lying ones, we have been able to identify the energetically four lowest-lying isomers.

Our method combines randomness with regularity and is significantly more 
efficient compared to the case when only random starting  
geometries are used. In addition, it is completely unbiased. However, as for any
structure-determination method, there is no guarantee that we have identified 
the true global total-energy minima, although we believe that we in most cases
have very good candidates. But, unfortunately, in two cases, $N=75$ and $N=104$
(both cases correspond to decahedral structures, which often have been difficult
to find in unbiased structure optimizations), our approach
failed and we had to include that in our study explicitly, and in another, $N=38$,
we only got the previously obtained structure after considerably more attempts than
otherwise used. We add that similar problems have been observed and analysed in studies on
Lennard-Jones clusters.\cite{wa97}

\section{Results and discussions}

\subsection{Small clusters --- accessing the accuracy}

The smallest clusters, i.e., Ni$_N$ for $2\le N\le 13$, provide an excellent 
test for our approach of two reasons: The clusters are so small that the number
of (meta-)stable structures is small, which in turn means that the global 
total-minimum within a given approach can be identified. And since the EAM 
has its foundations in considerations for infinite, extended metallic systems
with largely delocalized electrons, the small clusters should be those that
with most difficulties can be treated by our approach. Therefore, we have in 
Table \ref{tab00} tabulated our obtained structural parameters for $2\le N\le 8$ and
$N=13$ in comparison with results of other theoretical or experimental studies.

The table shows that the structure, here given through the average bond length $\langle 
r\rangle$, is most often given within a couple of \% of the results of experiment or of
parameter-free electronic-structure calculations. Moreover, none of the other 
parameterized methods, i.e., the tight-binding approach or model potentials of
the Sutton-Chen, Gupta, or effective-medium type, gives results that are systematically
more accurate than our approach does. This is important since for the larger clusters
that we shall discuss here it is not possible to apply accurate parameter-free 
methods when simultaneously attempting to optimize the structure. 

The table also shows that the relative ordering of the different isomers may 
differ for different methods. Hence, for $N=7$ most parameterized methods predict
the D$_{5h}$ isomer to have a lower total energy than the C$_{3v}$ structure, whereas experiment
and parameter-free theoretical studies predict a reversed ordering. Therefore, it can
be important to calculate more different, energetically low-lying isomers when 
studying theoretically the properties of not very small clusters --- identifying 
a different isomer as the one of the lowest total energy may suggest a significant
disagreement between different studies. 

\begin{figure}[tbp]
\begin{picture}(18,17)
\put(1,0){\psfig{file=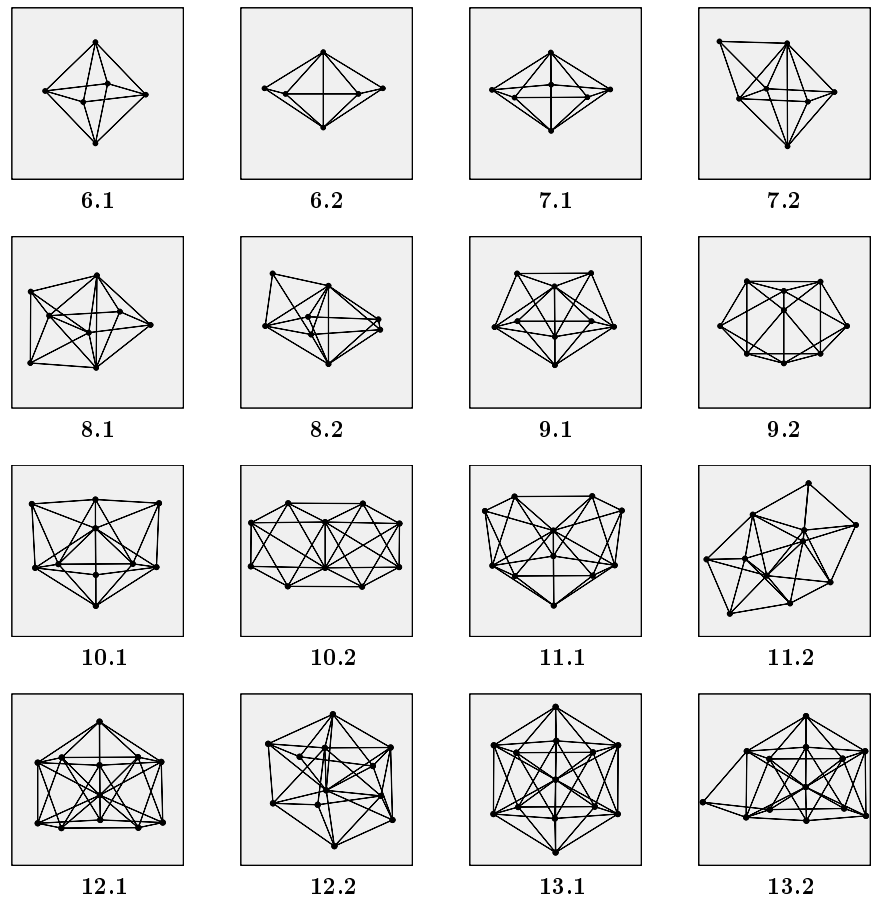,width=16cm}}
\end{picture}
\caption{The optimized geometries of the two energetically lowest isomers 
of Ni$_N$ clusters for $6 \le N \le 13$}
\label{fig01}
\end{figure}

In order to emphasize this point further we show in Fig.\ \ref{fig01} the structures
of the two energetically lowest isomers for $6\le N\le 13$. It is very clearly seen that
in some cases the two isomers are indeed very different and identifying the `wrong' one
as the most stable one may give unexpected results. 

It may also surprise that for the 
two isomers for $N=7$ that look very different we find a total-energy difference of
only 0.03 eV/atom. To our knowledge there exist two 
DFT calculations of Ni$_7$ cluster.\cite{de98,na96} Both studies predict for 
the two energetically lowest isomers a capped octahedron and a pentagonal 
bipyramid, respectively. The corresponding average binding energy per atom are
found to differ by only 0.05--0.07 eV/atom, i.e., very similar to our findings.

Table \ref{tab00} and Fig.\ \ref{fig01} give one further result for $N=7$: Ni$_7$ is
the smallest cluster that incorporates the fivefold symmetry axis, an important
element of icosahedral and decahedral symmetries. 

For $N=8$ our two energetically lowest isomers differ in energy by 0.016 eV/atom, 
in good agreement with the value of 0.02 eV/atom found by 
Desmarais {\it et al.} in a DFT study.\cite{de98}
There are some fundamental differences between 
different {\it ab initio} studies on Ni$_8$ cluster.\cite{de98,mi01,kr00,re95}
According to one  study \cite{re95} a cube  is the most stable 
structure for Ni$_8$, whereas in other studies\cite{de98,kr00} a cube was only the 
5{\it th} isomer. Further, in another work\cite{kr00} the second 
isomer was found to be a bicapped trigonal prism (C$_{\rm 2v}$) and a capped 
pentagonal bipyramid was only isomer number four. Michelini 
{\it et al.}\cite{mi01} 
has found for the eight-atom nickel cluster only one isomer with the tetrahedral 
symmetry T$_d$. This GGA result is inconsistent with all other studies. 
It seems that the LSDA calculations of Desmarais {\it et al.}\cite{de98} have the 
best agreement with both experimental and semiempirical studies. 

To our best knowledge there are only DFT studies\cite{ca99,re95} on Ni$_N$, 
$9 \le N \le 13$, for $N=13$. At $N=13$ the first Mackay icosahedron (I$_h$) is constructed.
The second isomer is 1.16 eV less stable. This is a rather large value 
and the icosahedron as the most stable structure for Ni$_{13}$ was found 
almost in all theoretical studies (the only exception being an extended-H\"uckel
study\cite{cu98}).
Once again we obtain good structural agreement with {\it ab initio} 
studies.\cite{ca99,re95} The obtained bond lengths of 2.36 (center to vertex) 
and 2.48 {\AA} (vertex to vertex) agree well with those derived by Calleja
{\it et al.},\cite{ca99} i.e., 2.41$\pm 0.03$ and 2.53$\pm 0.03$ {\AA},  
and with those of Reuse and Khanna,\cite{re95}, i.e., 
2.23 and 2.34 {\AA}, respectively. Furthermore, when comparing the differences in 
the binding energies per atom between the Ni$_{13}$ and Ni$_{2}$ clusters, 
our value of 1.525 eV/atom agrees very well with the corresponding value of 
1.558 eV/atom obtained using the self-consistent DFT 
program SIESTA.\cite{ca99} 

Closing this subsection we emphasize that for our purpose, the calculation
of energetic and structural properties of Ni$_N$ clusters, the present EAM approach
appears to be sufficiently accurate. 

\subsection{Energetic properties}

\begin{figure}[tbp]
\begin{picture}(16,10)
\put(0,0){\psfig{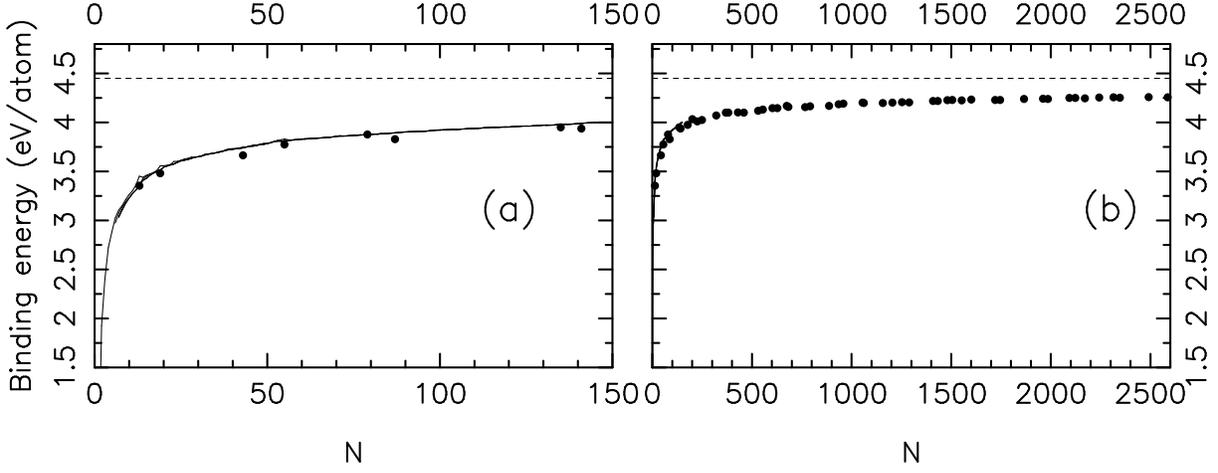}}
\end{picture}
\caption{Binding energy per atom as a function of size for the four energetically
lowest isomers for $N$ up to 150 (continuous curves) together with that of relaxed, 
spherical parts of the {\it fcc} crystal structure with an atom at the center 
(closed circles). The horizontal dashed
lines mark the value for bulk Ni as obtained with the same method.}
\label{fig02}
\end{figure}

In Fig.\ \ref{fig02} we show the binding energy per atom for the optimized clusters
for the four energetically lowest isomers together with the binding energy per atom
for clusters that were constructed as relaxed, spherical parts of the {\it fcc} crystal structure
(with an atom at the center of the sphere) for $N$ up to almost 2600. With few expectations
the four total-energy curves are so close that their difference can not be resolved in
the figure. On the other hand, except for $N=79$, the {\it fcc}-derived clusters lie
all below the other structures, indicating that for $N$ up to around 150 the {\it fcc} structure
is not playing a significant role as a structural motif (an exception occurs for $N=38$ for a
fragment of the {\it fcc} crystal with the center not at an atom --- this case is not
shown in the figure). Fig.\ \ref{fig02}(b) shows 
that the apparent saturation is not real: first when considering considerably larger
clusters, the convergence towards an average binding energy around 4.45 eV/atom can be
recognized.

\begin{figure}[tbp]
\begin{picture}(15,10)
\put(0,0){\psfig{file=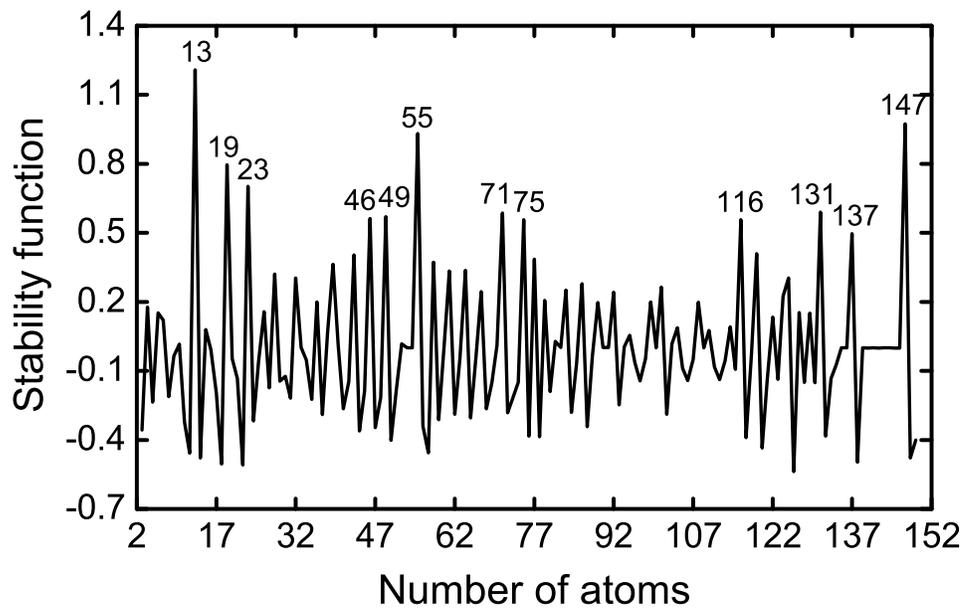,width=14cm}}
\end{picture}
\caption{The stability function as a function of cluster size.}
\label{fig03}
\end{figure}

In order to identify the particularly stable clusters we shall consider two criteria.
First, a cluster may be considered very stable if its binding energy per atom is much larger than
that of the two neighboring clusters. This can be quantified through the stability function,
$E_{\rm tot}(N+1.1)+E_{\rm tot}(N-1.1)-2E_{\rm tot}(N.1)$, where $E_{\rm tot}(N.k)$ is the
total energy of the energetically $k$-lowest isomer of the Ni$_N$ cluster. This function, 
that has maxima for particularly stable clusters, is shown in Fig.\ \ref{fig03}. Here 
we can identify a large number of particularly stable clusters, i.e., so-called magic
clusters. These are found for 
$N$ = 4, 6, 10, 13, 15, 19, 23, 26, 28, 32, 36, 39, 43, 46, 49, 55, 58, 61, 
64, 67, 71, 75, 77, 79, 83, 86, 89, 92, 95, 99, 101, 104, 108, 110, 114, 
116, 119, 122, 125, 127, 129, 131, 137, and 147. The most pronounced peaks occur at 
$N$ = 13, 19, 23, 46, 49, 55, 71, 75, 116, 131, 137, and 147 and most of them contain 
an odd number of atoms and, as seen in Table \ref{I}, these clusters 
possess high symmetry, including the highest icosahedral symmetry for Ni$_{13}$, Ni$_{55}$, and 
Ni$_{147}$.

\begin{figure}[tbp]
\begin{picture}(15,20)
\put(0,0){\psfig{file=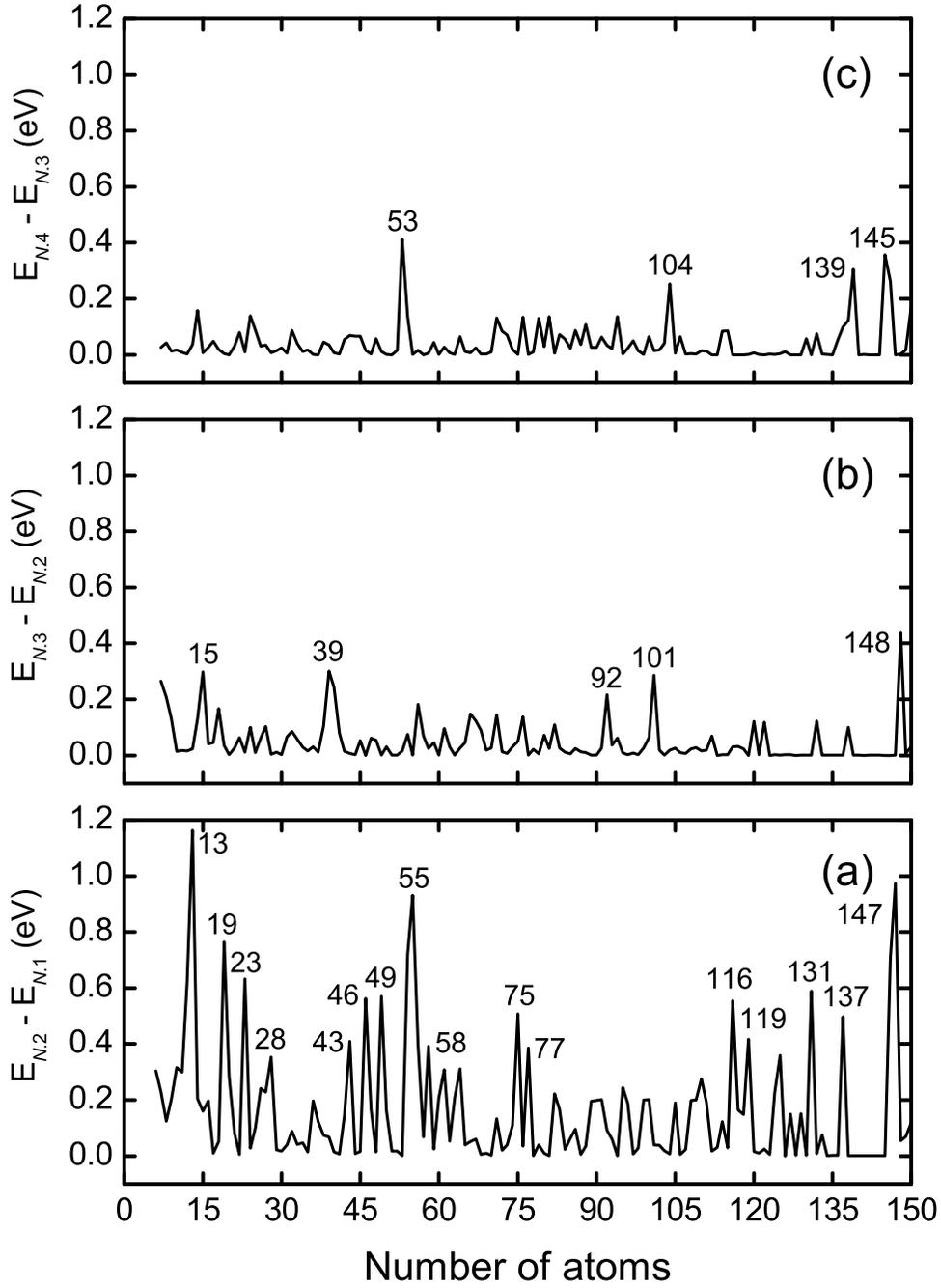,width=14cm}}
\end{picture}
\caption{The total-energy difference between the two energetically neighboring 
isomers as a function of cluster size. (a) shows the difference between the second 
and the first, (b) between the third and the second and (c) between fourth and 
the third isomer.}
\label{fig04}
\end{figure}

According to our other criterion for a particularly stable cluster, such a cluster occurs
if the energy difference between the two energetically lowest isomers $E_{\rm tot}(N.2)-
E_{\rm tot}(N.1)$ is large. This energy difference together with the energy differences
to the energetically higher ones are shown in Fig.\ \ref{fig04}. It is striking that the
energetically higher-lying isomers become energetically less and less separated and, as
seen when comparing with Fig.\ \ref{fig03}, many of the clusters that are particularly 
stable according to the first criterion is it also according to the second one. 

\subsection{Structural properties}

A central issue of our work is to identify general properties of the Ni$_N$ clusters from 
the large amount of information that we have obtained from the calculations. Instead of
discussing the individual clusters, in particular their structure (a brief summary of the 
structures is given in the Appendix), we shall introduce different
quantities that are devised to reduce the available information to some few key numbers.

\begin{figure}[tbp]
\begin{picture}(15,18)
\put(0,0){\psfig{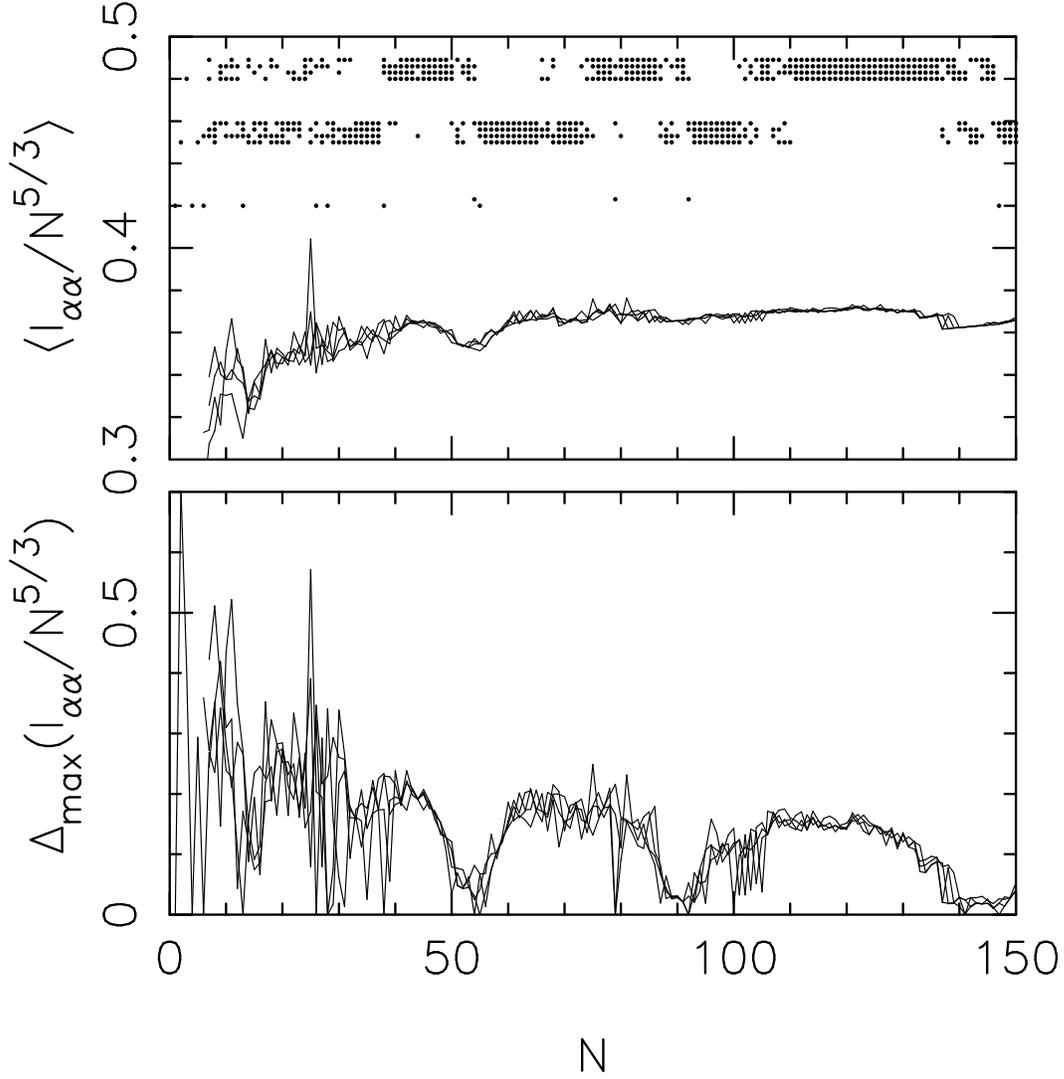}}
\end{picture}
\caption{Different properties related to the eigenvalues $I_{\alpha\alpha}$. In the upper panel
we show the average value together with points indicating whether clusters with overall
spherical shape (lowest set of rows), overall cigar shape (middle set of rows), or overall lens 
shape (upper set of 
rows) are found for a certain size. Moreover, in each set of rows, the lowest row corresponds
to the energetically lowest isomer, the second one to the energetically second-lowest
isomer, etc. In the lower panel we show the maximum difference of the
eigenvalues for the four different isomers.}
\label{fig05}
\end{figure}

First we shall consider the overall shape of the clusters. As we discussed in our earlier
report,\cite{cpl03} it is useful to examine the $3\times 3$ matrix containing the elements
\begin{equation} 
I_{st}=\frac{1}{u_l^2}\sum_{n=1}^N (R_{n,s}-R_{0,s})(R_{n,t}-R_{0,t})
\end{equation}
with $u_l$ = 1\AA\, being our length unit, and 
$s$ and $t$ being $x$, $y$, and $z$, and with 
\begin{equation}
\vec R_0=\frac{1}{N}\sum_{n=1}^N \vec R_n
\end{equation} 
being the center of the cluster. The three eigenvalues of this matrix, 
$I_{\alpha\alpha}$, can be used in separating the clusters into being overall spherical
(all eigenvalues are identical), more cigar-like shaped (one eigenvalue is large, the other
two are small), or more lens-shaped (two large and one small eigenvalue). Moreover, the
average of the three eigenvalues, $\langle I_{\alpha\alpha}\rangle$, is a measure of the
overall extension of the cluster. Finally, for a homogeneous sphere with $N$ atoms, the 
eigenvalues scale like $N^{5/3}$. Therefore, we show in Fig.\ \ref{fig05} quantities
related to these eigenvalues but scaled by  $N^{-5/3}$.

In the figure we see that only some few clusters have an overall spherical shape (these are
found for the energetically lowest isomer for $N$ = 4, 6, 13, 26, 28, 38, 55, and 147 and
for the next one for $N$ = 54, 79, and 92), which all correspond to high-symmetry isomers
(cf.\ Table \ref{I}) and, for the lowest-energy isomer, also to the class of magic clusters. 
Moreover, it is interesting to notice that the average value follows
more or less the same curve for all four isomers, i.e., the more compact ones (with relative
low average value) are found for $N$ just above 50 and above 140). Also the
largest differences follow more or less the same curve, except for some few cases mainly for
$N$ below 40. Therefore, except when the eigenvalues all are very similar (i.e., the largest
difference is very small, which occurs for $N$ around 50, 100, and 140), the overall shape
(i.e., lens- or cigar-like) is the same for all four isomers.

In our earlier work\cite{cpl03} we showed that the structural development of the isomer of
the lowest total energy with advantage could be monitored through so-called similarity
functions. Thus, for a given cluster Ni$_N$ we define a radial distance for each of the
atoms
\begin{equation}
r_n=\vert\vec R_n-\vec R_0\vert
\end{equation}
and sort these in increasing order.

\begin{figure}[tbp]
\begin{picture}(15,20)
\put(2,0){\psfig{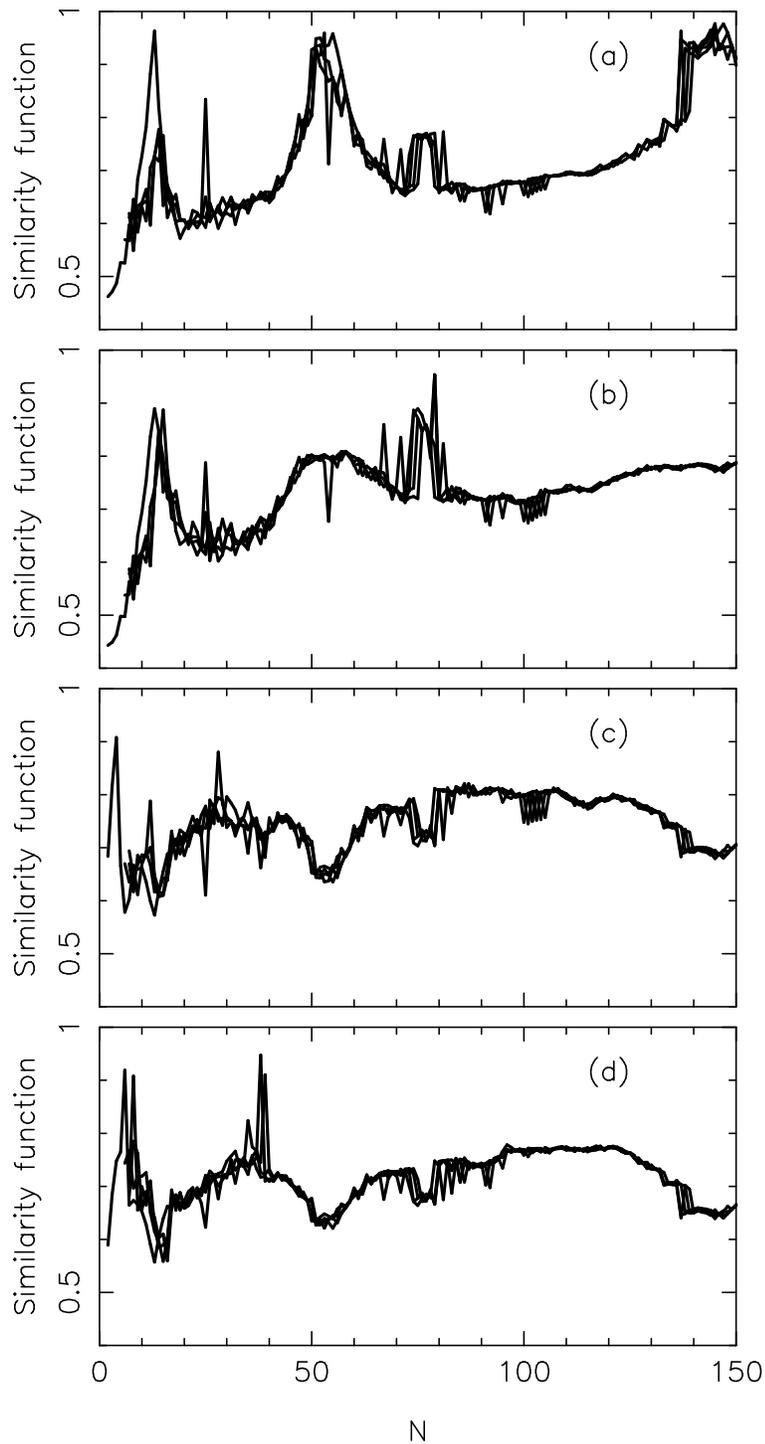}}
\end{picture}
\caption{Each panel shows the similarity function for all four isomers when comparing with
(a) an icosahedral cluster, and (b--d) a spherical fragment of the {\it fcc} crystal when the 
center of the fragment is placed at (b) the position
of an atom, (c) the middle of a nearest-neighbor bond, and (d) the center of the cube, 
respectively.}
\label{fig11}
\end{figure}

Simultaneously we consider a large fragment of a {\it fcc} crystal or, alternatively, a large
cluster of icosahedral symmetry. Also for these we define a radial distance for each atom,
$r_n'$, that also are sorted. In order to compare the given cluster with any of those systems
we calculate subsequently
\begin{equation}
q=\bigg[\frac{1}{N}\sum_{n=1}^N (r_n-r_n')^2\bigg]^{1/2},
\end{equation}
giving the similarity function
\begin{equation}
S=\frac{1}{1+q/u_l}
\end{equation}
($u_l$ = 1\AA),
which approaches 1 if the Ni$_N$ cluster is very similar to the reference system.
In Fig.\ \ref{fig11} we show the resulting functions in four cases, i.e., when comparing
with the relaxed Ni$_{309}$ cluster of icosahedral symmetry, and when comparing with
three fragments of the {\it fcc} crystal differing in the position of the center (the position
of an atom, the middle of a nearest-neighbor bond, and the center of the cube, respectively).

We see that both clusters that clearly resemble {\it fcc} fragments and such that resemble
icosahedral clusters can be identified. Of obvious reasons, the latter is the case
for $N$ around 13, 55, and 147, but it is interesting to notice that these clusters are
not singular: clusters with a large similarity to icosahedral clusters are found over a
larger range of clusters ($\sim$10) around the truly icosahedral cluster. Similarly, 
clusters that are close to being fragments of the {\it fcc} crystal can be identified 
around $N$ = 13, 28, 38, and 79. Moreover, it is surprising to observe that all four 
isomers for a given size show very similar similarity functions. 

\begin{figure}[tbp]
\begin{picture}(15,18)
\put(2,0){\psfig{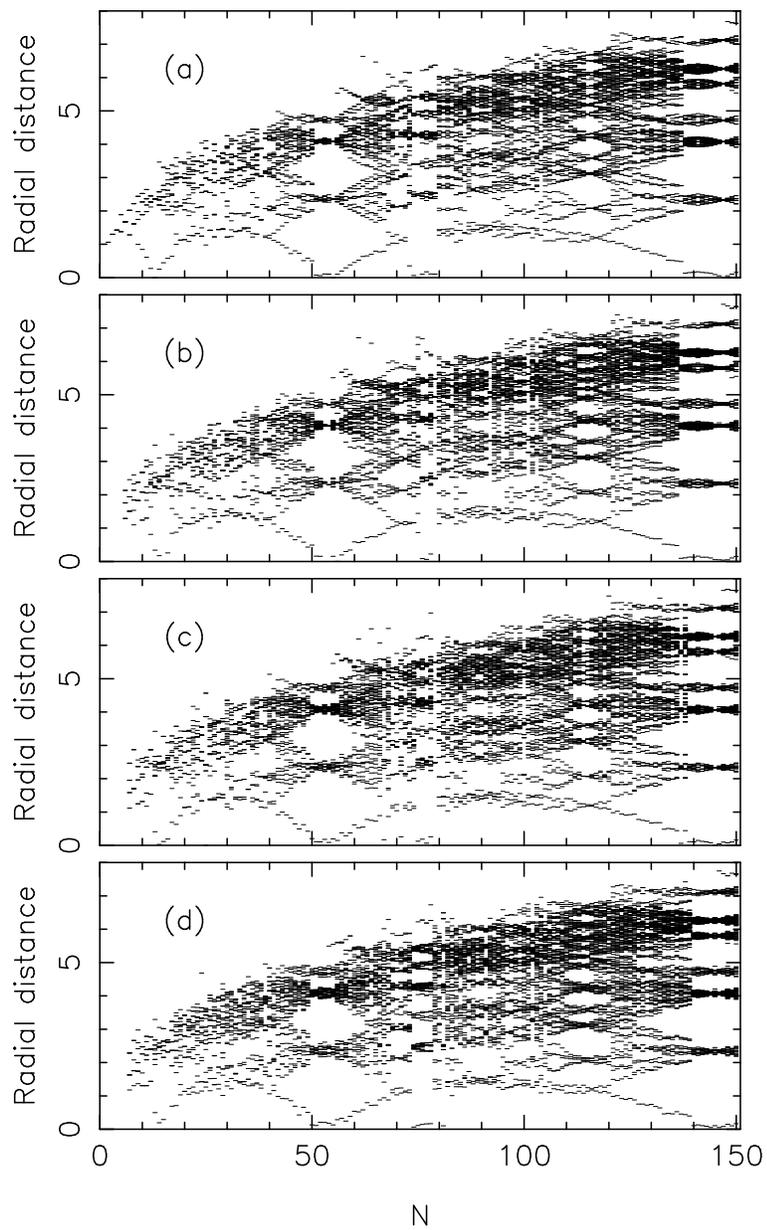}}
\end{picture}
\caption{Each panel shows the radial distances (in \AA) as a function of cluster size, i.e., each
small line represents (at least) one atom with that radial distance. The energy of the
isomers increases from the top to the bottom.} 
\label{fig10}
\end{figure}

Also the radial distances themselves can give some interesting information. These are shown
in Fig.\ \ref{fig10} for the four different isomers as functions of cluster size. Up to $N$
around 40, only few trends can be identified (with the exception of $N=13$), but for 
$N$ just above 50 a clear tendency towards shell-construction can be seen. Whereas this 
transition towards shell-construction is fairly smooth as a function of cluster size, 
independent of whether we start from the larger or the smaller clusters, in the range $75\le N
\le 80$ abrupt changes occur also towards some shell-construction. That this size range 
is special shall be discussed in more detail below. Also for $N$ close to 115 this tendency
is seen, but most dramatic are the results for clusters around 140 and above where, for
all four isomers, a very clear shell-construction is identified. Comparing with Fig.\ \ref{fig11}
we see that this is correlated to the construction of icosahedral-like structures. Thus, since
the infinite crystal possesses the {\it fcc} structure, the clear shell-construction in this
size range will have to disappear again for larger clusters.

In total, it can not surprise that when analysing the
structures of the individual clusters in detail we can identify many structural motifs
that are characteristic for clusters of particularly high stability also for clusters with
somewhat different sizes. Moreover, such an analysis\cite{unpub} shows also that many of
our structures for the smaller $N$ (these are the only ones that have been the subject of
earlier unbiased structure-optimizations) are close to those that have been
found in earlier studies. Here, however, we shall largely omit the discussion of the individual
structures but instead explore the more general trends.

\begin{figure}[tbp]
\begin{picture}(15,20)
\put(0,0){\psfig{file=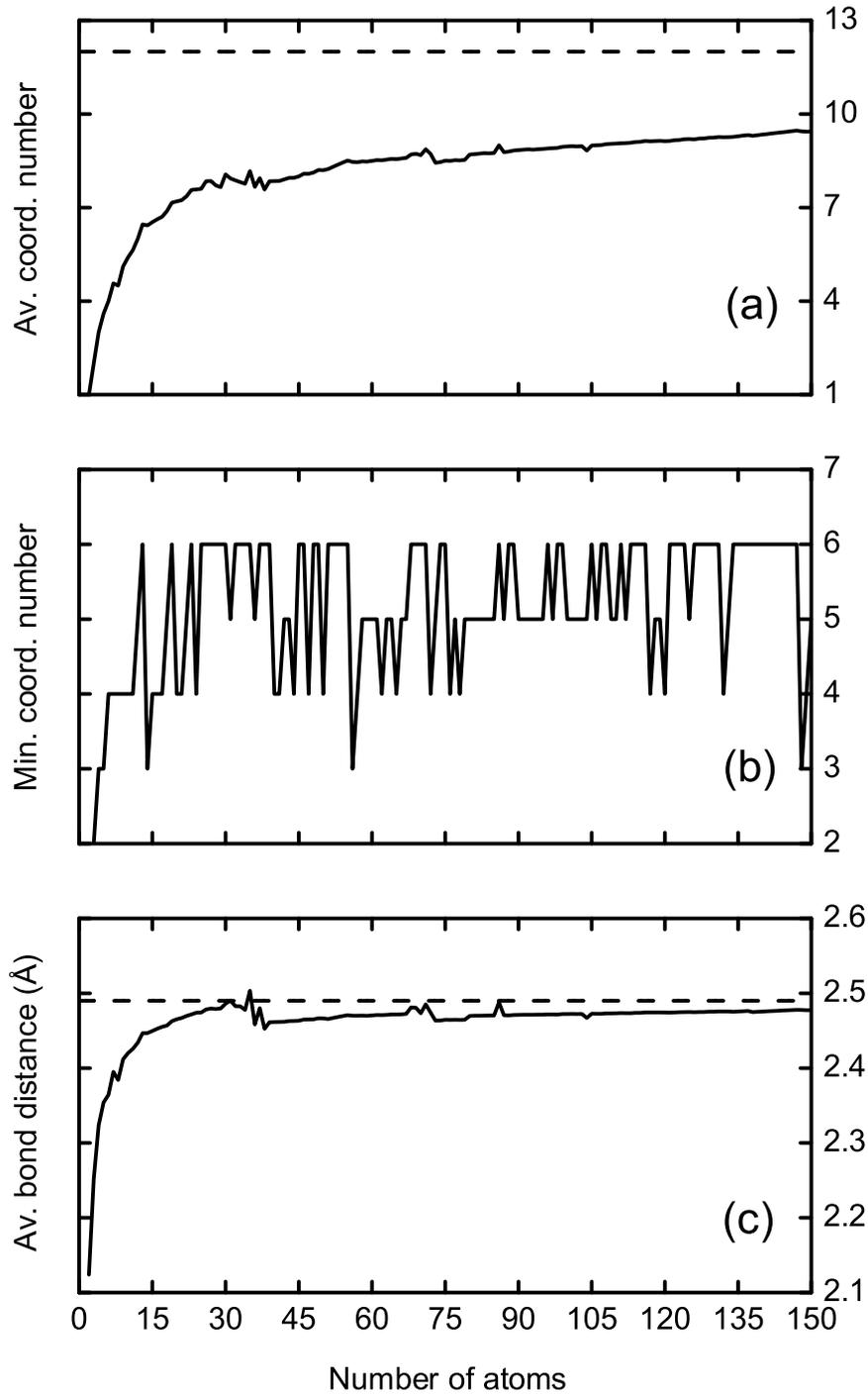,width=14cm}}
\end{picture}
\caption{(a) the average coordination number, (b) 
the minimum coordination number, and (c) the average bond distances as 
functions of cluster size. The dashed lines in (a) and (c) show the 
corresponding bulk values for nickel.}
\label{fig06}
\end{figure}

In Fig.\ \ref{fig06} we show some few quantities related to the structure of the 
clusters, i.e., coordination numbers and average bond lengths. Here, we define two
atoms as being bonded if their interatomic distance is less than 3.00 \AA, 
which is the average of the nearest-neighbor distance  (2.49\ {\AA}) and 
the next-nearest-neighbor distance (3.52\ {\AA})  in bulk Ni. Moreover, 
we distinguish between inner atoms with a coordination number of 12 or larger and surface atoms 
with a coordination number less than 12. 

Fig.\ \ref{fig06}(a) shows the average coordination number as function of $N$, where an
obvious increase as function of size is observed, but with additional oscillations
in particular for the smallest clusters. This function shows clearly a saturation towards
the bulk limit of 12, although one has to remember that even for the largest cluster
of our study 94 out of 150 atoms are characterized as surface atoms.

Fig.\ \ref{fig06}(b) shows the minimum atomic coordination for each cluster
size from the range $N=2-150$, which is useful information for chemical-probe
experiments due to the enhanced chemical activity of clusters containing 
low-coordination atoms. In addition, 
the existence of low-coordination atoms (i.e., coordinations of 3 or 4) could 
point to the occurrence of a cluster growth, where extra atoms are added to the surface of
the cluster, whereas the higher minimum-atomic-coordination
numbers ({\it e.g.} 5 or 6) could indicate a growth  where atoms are inserted inside the 
cluster, or, alternatively, upon a strong rearrangement of the surface atoms. These 
considerations follow the 
so-called "maximizing the minimum coordination".\cite{st92,we96} 

Fig.\ \ref{fig06}(c) shows the average bond length
as function of cluster size. The dashed line corresponds to the bulk 
value of 2.49\ {\AA}. The average bond length for all the structures 
with the exception of Ni$_{31}$, Ni$_{35}$, Ni$_{71}$, 
and Ni$_{86}$ is smaller than the bulk value. Such a reduction is in accordance 
with the results of most other theoretical studies, but it is interesting to observe
that this property approaches the bulk value faster than the other quantities of 
the figure.

Important experimental work in the determination of the structure of individual 
Ni$_N$ clusters has been presented by Riley $et$ $al.$\cite{pa94,pa95,pa95a,pa97,pa98,pa01} 
using chemical-probe experiments. Thereby, the number of N$_2$ molecules that is bonded
to a given cluster is determined, and in order to relate this number to the structure
of the cluster, the following empirical rules have been formulated:\cite{pa94,pa95,pa95a} 
$(i)$ nickel atoms with a coordination of four and less will bind two 
nitrogen molecules at saturation, $(ii)$ nickel atoms with coordination 
of five to eight will bind only one nitrogen molecule at saturation, 
$(iii)$ nickel atoms with a coordination of nine will bind one nitrogen 
molecule weakly or not at all (this kind of binding has not been 
observed\cite{pa95a} for $N>49$), and $(iv)$ nickel 
atoms with a coordination more than nine do not bind any nitrogen molecules. These rules
allow us in turn to estimate the number of N$_2$ molecules that will bond to a given 
cluster. We stress that the chemical-probe experiments provide information on the 
complete system, i.e., cluster plus adsorbants, whereas we consider the naked 
clusters. Although this difference may result in some deviations between theory and
experiment, a comparison should give useful information on the clusters.  This is
supported by the studies of Parks {\it et al.}\cite{pa94,pa95} that show that in
many cases the structures of naked and N$_2$-covered clusters are very similar.

\begin{figure}[tbp]
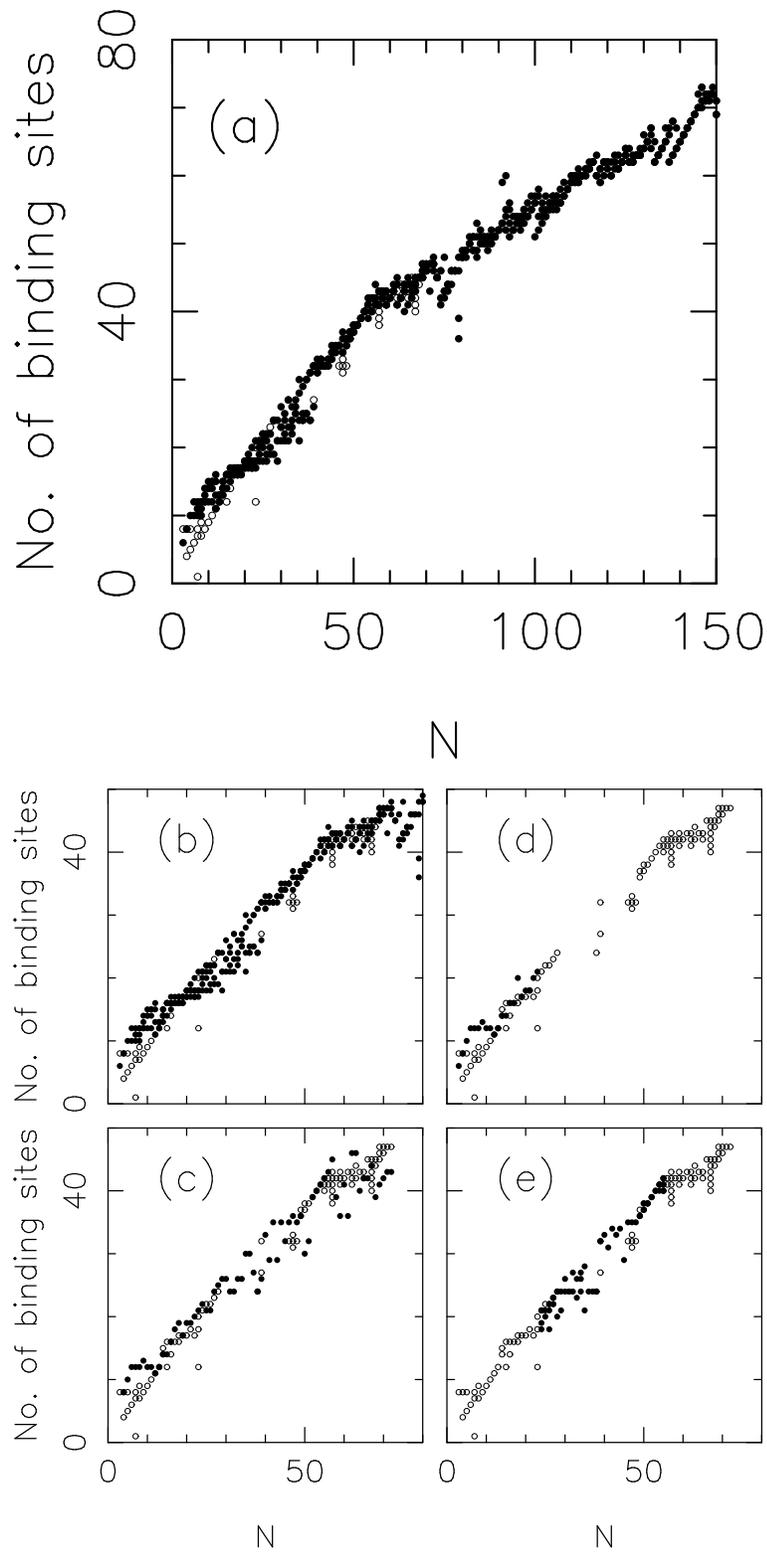

\begin{picture}(15,20.5)
\put(2,10.5){\psfig{file=fig07a.ps,width=10cm}}
\put(2,0){\psfig{file=fig07bcde.ps,width=10cm}}
\end{picture}
\caption{The (open circles) experimentally determined number of N$_2$ binding sites
for Ni$_N$ clusters from Refs.\ [\onlinecite{pa97,pa98,pa01}] as 
function of $N$ in comparison with 
theoretical calculated numbers (closed circles) from (a,b) the present work, (c) 
from [\onlinecite{do98}], (d) from [\onlinecite{na97}], and (e) from [\onlinecite{we96}].}
\label{fig07}
\end{figure}

Doing so, we obtain the results of Fig.\ \ref{fig07}, where we also compare with the results of
other theoretical works. The experimental work covers the range $3\le N\le 72$, whereas the
former theoretical works cover $4\le N\le 80$ (Ref.\ [\onlinecite{do98}]), $3\le N\le 
23$ (Ref.\ [\onlinecite{na97}]), and $24\le N\le 55$ (Ref.\ [\onlinecite{we96}]), obtained
using the Sutton-Chen potential,\cite{do98} the Finnis-Sinclair potential,\cite{na97}
and the effective-medium theory.\cite{we96} Fig.\ \ref{fig07}(a), where we show the
results for the whole range of our study, i.e., $2\le N\le150$, shows an interesting 
aspect that obviously is beyond most of the previous experimental and theoretical
studies, i.e., that up to around $N\simeq 55$ the number of binding sites increases
more rapidly with $N$ than for $N$ above $\sim$55, i.e., the curve is roughly two
piecewise linear curves. The experimental studies [cf.\ Figs.\ \ref{fig07}(d) and (e)]
do in fact show a small tendency in this direction, which, on the other hand, is not
found in the earlier theoretical work [cf.\ Fig.\ \ref{fig07}(c)]. Figs.\ \ref{fig07}(b)-(e)
show a comparison of the experimental results with those of the four different 
theoretical studies to this issue. It is clear that our approach is at least as accurate
as any of the other theoretical ones. Moreover, the Sutton-Chen potential seems to
predict a larger scatter of the number of binding sites as a function of size compared
to any other approach. 

Whereas icosahedral motifs are found to play an important role for very many values of $N$,
we found that for $N$ around 75 also other structural motifs are important. The chemical-probe
results for N$_2$ uptake, both the experimental ones (notice, however, that the experimental 
studies are limited to $N\le 72$) and the simulated, theoretical ones, suggest also that
in this range of $N$ values more structural motifs compete: for these values of $N$ the
results of Fig.\ \ref{fig07} are particularly scattered. Moreover, as we shall see in the
next subsection, also growth processes suggest that these $N$ values are special. Therefore,
we shall study the structural properties of those clusters in some more detail.

\begin{figure}[tbp]
\begin{picture}(15,13)
\put(0,0){\psfig{file=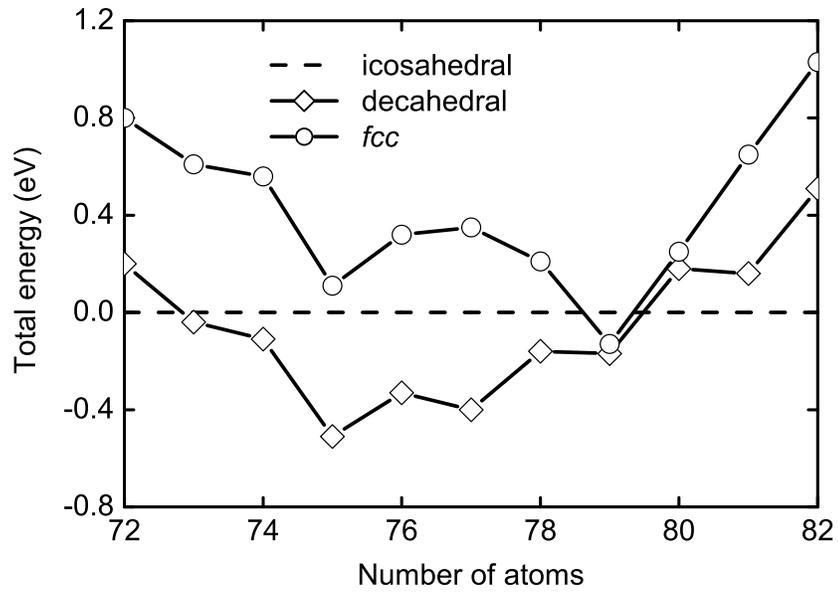,width=14cm}}
\end{picture}
\caption{The relative stability of the icosahedral, decahedral and 
{\it fcc} structures as function of cluster size for $72 \le N \le 82$.}
\label{fig08}
\end{figure}

For cluster sizes ranging between 70 and 80 atoms two more closed-shell structures
other than icosahedral ones can be identified, i.e., the (2,2,2) Marks decahedron 
(giving a 75-atom cluster) and a 79-atom {\it fcc} cluster, which is the truncated centered
octahedron. Due to their quasi-spherical shape they have a particularly low surface energy. 
Moreover, the cluster sizes between  70 and 80 are very different from the values of 
closed-shell icosahedral clusters that have 55 or 147 atoms. Accordingly, one may expect 
a strong competition between these three main morphologies in this size range.
Our calculations support this suggestion.  
Fig.\ \ref{fig08} shows the lowest total energies for the decahedral 
and {\it fcc} clusters relative to the 
icosahedral ones for $72 \le N \le  82$. The Marks decahedron and its derivatives 
dominate over the icosahedral structures for $73 \le N \le 79$. At $N=79$ 
both decahedral and {\it fcc} motifs are more stable than the 
icosahedral one. The decahedron at $N=79$ is only 0.04 eV more stable than 
the truncated octahedron. Finally, three of four lowest-energy isomers at $N=75$ are 
decahedra and for $N=76-78$ there are only decahedral clusters among the four first isomers.

This result suggests that a fundamental structural change takes place for $N$ around 75.
This, however, is only partly the case: for larger $N$ values the icosahedral packing is 
again stabilized (an exception occurs for $N$ around 104, where decahedral structures
are found). In order to verify this we computed the total energies
for several optimized closed {\it fcc} and decahedral structures, i.e., truncated noncentered
octahedra for $N=116$ and 140, a noncentered octahedron and a (3,2,2) Marks decahedron 
for $N=146$, and a cuboctahedron and an Ino decahedron for $N=147$. In all cases 
the icosahedral structures were found to have lower total energies. 
The {\it fcc} clusters for $N=116$, 140 
and 146 and {\it m}-Dh$_{146}$ were 0.85, 2.01, 5.64 and 2.29 eV, respectively, less stable than 
the icosahedral-like structures. The three-shell Mackay icosahedron 
was 5.69 and 4.71 eV more stable than the cuboctahedron and the Ino decahedron, 
accordingly. Thus, our calculations point strongly to icosahedral growth up to cluster
sizes of at least 150. This statement is supported by mass-spectroscopy
studies of nickel clusters\cite{pe94} which suggest icosahedral growth over a 
wide cluster-size region from 50 to 800. 

\subsection{Growth patterns}

One of the central issues is to identify how clusters grow, i.e., whether
some fundamental principles determine how (and whether) the cluster with $N$ atoms is derived 
from the one with $N-1$ atoms. We shall here study this issue from a completely static 
point of view, i.e., dynamic and kinetic effects, that most likely are important in 
experiment, will be neglected here. Basic information about this issue is given by the structures
of the magic clusters. The majority of the structures of these are icosahedral. Thus our study 
supports predominantly icosahedral growth of nickel clusters.
There are two different ways of icosahedral growth. Thus, 
the first Mackay icosahedron, Ih$_{13}$, contains 
20 triangular faces joined by 30 edges and 12 vertices. In the first case, 
MIC/Mackay or {\it fcc} covering,\cite{mo96,al00,do95,ha84,su94} additional 
surface atoms are placed on top of the edges and the vertices. This kind of growth 
results ultimately in the second Mackay icosahedron, Ih$_{55}$. Another possibility,
the TIC/polyicosahedral\cite{mo96,al00} or {\it hcp} or face-capping growth
(see Ref.\ [\onlinecite{do95}] and references therein), corresponds to adding
atoms on top of the atoms at the center of each face ($T$ sites) which leads, 
after complete covering, to the  45-atom rhombic tricontahedron. 
The TIC structures are the structures with the shorter average bond length, 
but at the same time they possess the larger strain energy. They are favored 
at the beginning of a covering until some critical size (depending on 
details of the atom-atom interaction\cite{ha84,su94}) after which the MIC 
structures become lower in energy. 

To explain the icosahedral shell filling in Mg and Ca clusters Martin 
{\it et al.}\cite{ma91} proposed the so-called {\it umbrella} model. 
According to this model enhanced stability is achieved each time a 
vertex and its five surrounding faces are covered (capping umbrellas). 
The capping umbrellas for MIC and TIC structures lead to different 
magic numbers. By covering of Ih$_{13}$ the MIC umbrellas are 
completed at $N=19$, 24, 28, 32, 36, 39, 43, 46, 49 and 55, whereas 
the TIC umbrellas give magic numbers for 19, 23, 26, 29, 32 and 34.
Comparing with our set of magic numbers (see above), 
we find that growth from Ni$_{13}$ to  Ni$_{27}$ is icosahedral and 
it can be described according to the umbrella model. For $N>27$ our 
magic numbers coincide with the MIC magic figures. This 
confirms seemingly the suggestion\cite{mo96} that at $N=27$ there is a  
transition from TIC to MIC growth. However, a detailed analysis of our 
structures shows that our 
structures for Ni$_{28.1}$ and Ni$_{32.1}$ are not icosahedral. Instead, the 
most stable icosahedral structure appears first at $N=35$ whereas the structures
according to the MIC growth or to  
{\it fcc} packing is found up to $N=56$ with a single exception for $N=38$.
Thus the transition from anti-Mackay to Mackay covering occurs in the region 
of cluster sizes from 28 to 34. Cluster growth in this range is complicated
and there is a competition between icosahedral, polyicosahedral and 
tetrahedral structure. Probably this non-trivial growth was responsible 
for the difficulties in interpreting results of nitrogen-uptake 
experiments\cite{pa94,pa95,pa95a} for $29 \le N \le 48$.

At $N=55$ we have the next complete icosahedron. 
According to the TIC growth model we should then have magic number at
$N=$58, 61, 64, 67,  and 71,\cite{mo96} which is in excellent agreement with our results.
At $N=71$ the TIC umbrella growth is completed.  
The MIC umbrella growth is completed 
at $N=71$, 83, 92, 101, 110, 116, 125, 131 and 137, again in agreement with our
calculated magic numbers.

Our study allows a more detailed analysis of possible growth processes that ultimately
will confirm the complexity of the growth. Thus, in order to quantify whether the cluster
of $N$ atoms is related to that of $N-1$ atoms we consider first the structure of the 
lowest total energy for the $(N-1)$-cluster. For this we calculate and sort all interatomic
distances, $d_i$, $i=1,2,\cdots,\frac{N(N-1)}{2}$. Subsequently we consider the $N$ fragments
of the $N$-cluster that can be obtained by removing one of the atoms and keeping the rest
at their positions. For each of those we also calculate and sort all interatomic distances
$d_i'$, and subsequently calculate 
\begin{equation}
q=\bigg[\frac{2}{N(N-1)}\sum_{i=1}^{N(N-1)/2} (d_i-d_i')^2\bigg]^{1/2}.
\end{equation}
Of the $N$ different values of $q$ we choose the smallest one, $q_{\rm min}$ and 
calculate the similarity function
\begin{equation}
S=\frac{1}{1+q_{\rm min}/u_l}
\end{equation}
($u_l$ = 1\AA) 
which approaches 1 if the Ni$_N$ cluster is very similar to the Ni$_{N-1}$ cluster plus
an extra atom. 

\begin{figure}[tbp]
\begin{picture}(15,20)
\put(2,0){\psfig{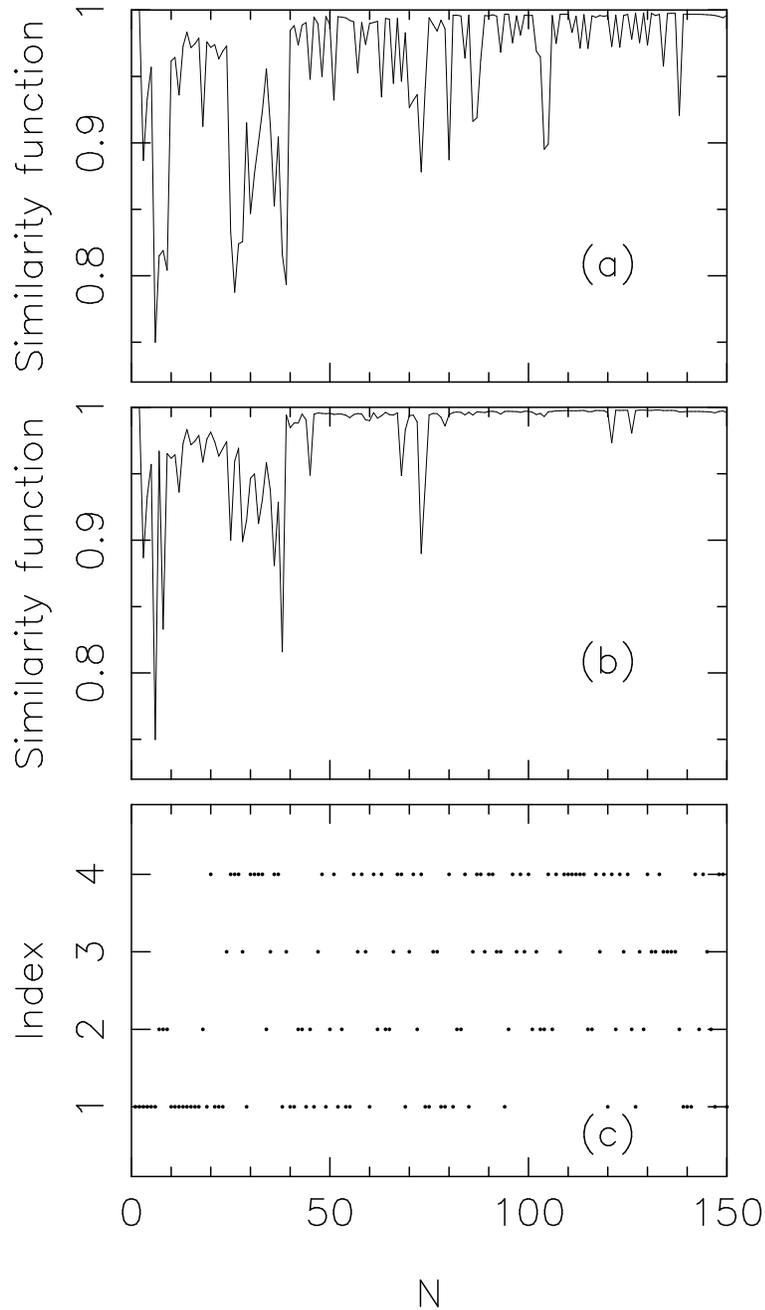}}
\end{picture}
\caption{(a) and (b) show the similarity functions that describe whether
the cluster with $N$ atoms is similar to that of $N-1$ atoms plus an extra 
atom, when (a) only considering the lowest-energy isomer for the $(N-1)$-atom
cluster and (b) considering all four isomers for that cluster. (c) shows
which isomer in the latter case is most similar to the one of $N$ atoms.}
\label{fig09}
\end{figure}

This function is shown in Fig.\ \ref{fig09}(a). We see indeed that for $N$ up to around 40,
$S$ is significantly different from 1, giving support for the consensus that in this range
the growth is complicated. For larger $N$ we see again that around $N\simeq75$ some larger
structural changes take place. Fig.\ \ref{fig03} shows, however, that the energetically lowest
isomers differ only little in energy. Therefore, one may ask whether 
isomers with a slightly higher energy 
play a role in the growth process. Accordingly, we examined whether the structure
of the energetically lowest isomer of Ni$_N$ would resemble $\underline{\rm any}$ 
of the four energetically lowest
isomers for Ni$_{N-1}$ in the same sense as above, i.e., we use the quantity $q$ in quantifying
the structural difference. Out of the four different values for $q$ (for the four different
isomers) we choose the smallest one and construct the similarity function from this. The 
resulting function is shown in Fig.\ \ref{fig09}(b), and compared with Fig.\ \ref{fig09}(a)
it is clearly seen that in particularly in the region $N>50$ the similarity function now 
approaches 1, except for the clusters around $N=70-75$, i.e., the region where we already
in analysing other properties have found atypical behaviors. In Fig.\ \ref{fig09}(c) we 
show which of the four energetically lowest isomers of Ni$_{N-1}$ leads to largest similarity
with the Ni$_N$ cluster, and from the fact that this only for the smaller $N$ most often 
is the first isomer, we learn that the growth process is a complicated process where more
different isomers are important, i.e., not only the energetically most stable one, although
these may resemble each other (cf.\ Fig.\ \ref{fig05}). This means that it is difficult to
imagine that the growth occurs as a one-by-one addition of atoms to the energetically most
stable isomers.

\section{conclusions}

In summary, we have determined the four energetically lowest isomers of nickel 
clusters Ni$_N$ in the range $2 \le N \le 150$, using a combination 
of the embedded-atom method in the version of Daw, Baskes and Foiles (for the calculation
of the total energy for a given structure), the 
{\it variable metric/quasi-Newton} method (for the determination of the closest total-energy
minimum), and our own {\it Aufbau/Abbau}
method (for the determination of the global total-energy minimum). Although the calculations
provide structural information for each individual cluster, separately,\cite{unpub} that may be 
useful information for experimental studies, we have refrained from discussing these 
separately (some information is, however, given in the Appendix), 
but instead focused on identifying general trends for Ni$_N$ clusters, and reducing
the information on the individual clusters to some few key quantities like total energy (per
atom), overall symmetry and shape, average bond length and coordination number, and similarity
with other structures ({\it fcc} fragments, icosahedral clusters, or smaller clusters).

Since our total-energy approach (the embedded-atom method) has its roots in arguments for
extended solids with delocalized atoms, we first verified that it could be applied also for
the fairly open structures of our interest. We found that for the absolutely most `critical'
cases, i.e., the smallest clusters, our results are in excellent agreement with available
theoretical and experimental information making us believe that our approach is accurate.
Furthermore, we have also tried to use the parameterization of Voter and Chen\cite{vo87,vo95}
of the EAM, which has been produced in order to give accurate descriptions of also finite
and more open systems, for some of the smallest clusters. To our satisfaction, no
structural changes were found and, therefore, we did not pursue this approach.

Our study predicts a number of particularly stable clusters, i.e., magic numbers, which in
many cases are in agreement with the prediction of other studies and in some cases with 
`common expectations' although we stress that in our study they were found in a completely
unbiased approach. These magic numbers were clearly visible both in the `stability function'
and in the total-energy difference between the energetically lowest and higher-lying isomers,
whereas for the energetically higher-lying isomers, the differences in the total energy become
small. 

We found also that even for large values of $N$ (above 2500) the total energy per atom has
still not converged to the bulk limit. Similarly, the average coordination number is for clusters
with $N$ around 150 still far from the bulk limit, whereas the average bond length has 
come close to the bulk limit for this size.

In order to study the structural properties we analysed the eigenvalues of the matrix 
containing the moments of inertia and introduced so-called similarity functions. The latter
indicated that all the four isomers we have studied here are very similar in most
cases, and that roughly
spherical clusters were found mainly for the energetically lowest isomer but in some cases 
also for the second-lowest one, and that these often correspond to particularly stable
structures. With the similarity functions we could identify clusters with a predominantly 
icosahedral structure and such with an essentially {\it fcc}-like structure. Whereas the 
former occurred for $N$ around 13, 55, and 147, the latter was found, e.g., for $N$ around
79. It was interesting to see that these structural motifs were built up over a larger 
range of $N$ values and were, accordingly, not limited to those singular values of $N$.

By analysing the distribution of radial distances as a function of cluster size we could
identify certain size regions, e.g., $N$ around 50, 75, 115, and above 140, where 
shell structures were clearly recognizable. This property, as well as several others,
showed, furthermore, that all four isomers for a given $N$ have very similar structures.

Chemical-probe experiments have been used in identifying the structure of the individual 
clusters, and for this we found an excellent agreement. Our study shows that for clusters
with $N$ around 50--60, a change occurs, so that above this size the number of N$_2$ binding
sites depends much weaker on $N$ than below this size. This change is just on the boarder of
the cluster sizes that have been studied experimentally and theoretically so far.

We used the magic numbers in analysing different models for cluster-growth processes.
Generally, the cluster growth is according to multilayered-icosahedral or layer-by-layer 
growth. We found that the one-, two-, and three-shell Mackay 
icosahedra are the most stable structures for 13-, 55-, and 147-atom nickel 
clusters (the latter for the first time in unbiased studies). This result agrees well with the 
suggestion in Ref.\ \onlinecite{cl91} that Mackay icosahedra are the most stable structures 
up to $N=2300$. But the cluster-growth pattern between two closed-shell 
icosahedra is not simple. It is predominantly icosahedral with islands of 
{\it fcc} growth for Ni$_6$, Ni$_{38}$, tetrahedral for Ni$_{28}$  and 
decahedral for Ni$_{73}$-N$_{79}$ and Ni$_{104}$. For $14 \le N \le 27$ the growth 
is polyicosahedral. Moreover, we found a growth according to 
{\it hcp}/anti-Mackay for $57 \le N \le 72$ 
and for $ 148 \le N \le 150$. From $N=35$ to $N=56$ (with the exception 
of $N=38$) and from $N=80$ to $N=147$ (except for $N=104$) 
the MIC cluster growth was found. Thus, the change-over from 
the {\it hcp}-type of cluster growth to the {\it fcc} one always happens 
within a transition region of cluster sizes and not immediately.

Moreover, whereas the growth process only with difficulties could be identified for
the smaller $N$ (due to the lack of similarity between the clusters), for larger $N$
we found that not only the energetically lowest isomers could be important in 
describing the growth. Thus, for all ranges of $N$, the growth is non-trivial.

A special case is that for $N$ around 75. More different properties suggested that
in this range special things occur. Analysing the total energy for different structures
in this size range we found indeed that more structures were energetically close and,
accordingly, would play a role for those clusters. Therefore, for these values of $N$
there is a rapid and dramatic change in structure as a function of $N$.

\begin{acknowledgments}
We gratefully acknowledge {\it Fonds der Chemischen Industrie} for very 
generous support.
 This work was supported by the SFB 277 of the University 
of Saarland. The authors are grateful to C.\ L. Cleveland, to Z.\ B. G\"uven\c{c} and 
M. B\"oy\"ukata, and to 
K. Michaelian and I.\ L. Garz\'on for providing 
us with the coordinates of some cluster structures. We would like to thank 
J.\ P.\ K. Doye for references to some recent studies on nickel clusters.
\end{acknowledgments}

\appendix

\section{Structure of the clusters}

In this appendix we shall present the structure of some of the clusters without entering
a detailed discussion of each clusters individually. We shall use the notation
$N.k$ for the Ni$_N$ cluster of the $k$-lowest total energy. We add that the nuclear
coordinates and total energies for the optimized clusters can be obtained from the 
authors upon request.

3.1: equilateral triangle. 4.1: tetrahedron. 5.1 trigonal bipyramid. 6.1 octahedron.
6.2: pentagonal bipyramid without a single pentagonal vertex. 7.1: pentagonal bipyramid.
7.2: capped octahedron. 8.1: bisdisphenoid. 8.2: capped pentagonal bipyramid.
9.1: bicapped pentagonal bipyramid. 9.2: tricapped trigonal prism. 10.1: tricapped pentagonal 
bipyramid. 10.2: trapezoidal antiprism. 11.1: quadricapped pentagonal bipyramid. 
11.2: two pentagonal bipyramids with a common face. 12.1: icosahedron without one atom at 
one vertex. 12.2: low symmetry. 12.3: trapezoidal bipyramid with capped trigonal faces. 
12.4: truncated trigonal bipyramid. 13.1: Mackay icosahedron. 13.2, 13.3, 13.4: similar
structures as 13.1, but with the move of a vertex atom to a face. 14.1: the structure of
13.1 with the capping of one of the trigonal faces. 14.2: similar, but with the 14th atom
on the top of a pentagonal edge. 14.3: bicapped hexagonal antiprism. 14.4: tricapped 
pentagonal prism. 15.1: bicapped icosahedron. 15.2: centered bicapped hexagonal antiprism.
15.3, 15.4: capped pentagonal prism. 

16.1: tricapped icosahedron. 18.1: double icosahedra without one atom in one pentagonal ring.
18.2: double icosahedra without one atom in a capping vertex. 19.1: double icosahedron.
19.2, 19.3, 19.4: derivatives of the double icosahedron by moving of an atom 
from an apex or from a pentagonal ring to a face-capped or a bridging position. 20.1: double 
icosahedron with an atom added to the a pentagon in a bridging position. 
21.1, 21.2, 21.3, 21.4: derivatives of the double icosahedron with two additional atoms
at bridging or capping positions. 22.1, 22.2: triple icosahedron without one atom 
either from a pentagonal ring (22.1) or from a capping position (22.2). 23.1: 
triple icosahedron. 24.1, 24.2: triple icosahedron with an additional atom. 25.1: 
three sequentially interpenetrating icosahedra whose vertices are located on one line.
26.1: six interpenetrating double icosahedra (a polyicosahedron). 27.1: the 
polyicosahedron plus an extra atom at a pentagonal cap. 28.1: tetrahedron.

In general, the structures for $29\le N\le 34$ do not display any dominant 
morphology --- there is a competition between tetrahedral, polyicosahedral,
and icosahedral packings. One can consider this size range as a transition range 
from {\it hcp} to {\it fcc} growth. 

29.1: distorted tetrahedron. 30.1: a polyicosahedral morphology. 32.1: derived from 
the 28.1 tetrahedron by adding four surface atoms.  35.1: MIC/Mackay structure. 36.1, 37.1:
contain icosahedral motifs. 38.1: truncated octahedron. 38.2, 38.3: icosahedral fragments.
39.1: the second Mackay icosahedron (55.1) without a 16-atom cap. 39.4: a {\it fcc} cluster.

49.1: the second Mackay icosahedron (55.1) without one face. 51.1, 52.1, 53.1, 54.1: obtained
from 49.1 by adding atom by atom until 55.1: second Mackay icosahedron. 54.2: as 55.1 but
without the central atom. 56.1: as 55.1 but with an extra atom at the center of a triangular
face. $N=56-71$: TIC/anti-Mackay growth in accordance with the umbrella model. 

75.1: (2,2,2) Marks decahedron. 79.2: a {\it fcc} cluster. 104.1: related (2,3,2) Marks decahedron.
147.1: three-layer icosahedron.

\begingroup
\squeezetable
\begin{table}
\caption{The symmetry (Sym.) and averaged bond length ($\langle r\rangle$) for smaller
Ni$_N$ clusters from different experimental and theoretical studies. EAM, TBMD, FS, SC, AI, EMT,
$n$G, and exp denotes Embedded-Atom calculations, tight-binding molecular-dynamics
calculations, Finnis-Sinclair potential, Sutton-Chen potential, {\it ab initio} calculations, 
effective-medium-theory calculations, many-body Gupta potential, and experiment, respectively.
For some values of $N$, more different meta-stable isomers have been found, and they have
then been characterized by an additional number (e.g., 7.1, 7.2, 7.3, and 7.4, with 7.1
being the stabler one, 7.2 the second-most stable one, a.s.o.)}
\begin{ruledtabular}
\begin{tabular}{ccccc|ccccc|ccccc}
$N$ &  Ref. & Sym. & $\langle r\rangle$ (\AA) & Method &
$N$ &  Ref. & Sym. & $\langle r\rangle$ (\AA) & Method &
$N$ &  Ref. & Sym. & $\langle r\rangle$ (\AA) & Method 
\\
\tableline
2 & here                    & D$_{\rm \infty h}$ & 2.12 & EAM & &  \onlinecite{do98} & O$_{\rm h}$        & -- & SC         &  &  \onlinecite{do98} & D$_{\rm 2d}$       & --   & SC      \\             
  &  \onlinecite{la96}  & D$_{\rm \infty h}$ & 2.20 & TBMD    & &  \onlinecite{lo94,mi99} & O$_{\rm h}$   & --  & $n$G      &  &  \onlinecite{lo94} & D$_{\rm 2d}$       & --   & $n$G    \\             
  &  \onlinecite{na97}  & D$_{\rm \infty h}$ & 2.01 & FS      & &\onlinecite{mi01}&C$_{\rm i}$ ($\sim$O$_{\rm h}$)&2.40&AI &  &  \onlinecite{mi01} & T$_{\rm d}$        & 2.32 & AI      \\             
  &  \onlinecite{mi01}  & D$_{\rm \infty h}$ & 2.13 & AI      & &  \onlinecite{re95} & D$_{\rm 4h}$       & 2.33 & AI       &  &  \onlinecite{re95}  & O$_{\rm h}$       & 2.16 & AI      \\             
  &  \onlinecite{re95}  & D$_{\rm \infty h}$ & 1.99 & AI      & &  \onlinecite{pa94}  & O$_{\rm h}$       & --   & exp      &  &  \onlinecite{de98} & D$_{\rm 2}$ ($\sim$D$_{\rm 2d}$) & 2.28 & AI \\   
  &  \onlinecite{ca99}  & D$_{\rm \infty h}$ & 2.17 & AI      & 6.2& here                   & C$_{\rm 2v}$     & 2.37 & EAM &  &  \onlinecite{kr00} &  D$_{\rm 2d}$      & 2.37 & AI      \\             
  &  \onlinecite{pi95}  & D$_{\rm \infty h}$ & 2.15 & exp     & &  \onlinecite{mi99} & C$_{\rm 2v}$       & --   & $n$G     &  &  \onlinecite{pa94} &  D$_{\rm 2d}$      & --   & exp     \\             
3 & here                    & D$_{\rm 3h}$       & 2.25 & EAM & &  \onlinecite{mi01} & D$_{\rm 2h}$       & 2.40 & AI       &  8.2&here                    & C$_{\rm s}$        & 2.40 & EAM \\          
  &  \onlinecite{la96}  & D$_{\rm 3h}$       & 2.30 & TBMD    & 7.1&here                    & D$_{\rm 5h}$     & 2.39 & EAM &  &  \onlinecite{bo_comm03}& C$_{\rm s}$       & 2.38 & EAM  \\             
  &  \onlinecite{na97}  & D$_{\rm 3h}$       & 2.10 & FS      & &  \onlinecite{la96} & D$_{\rm 5h}$       & 2.51 & TBMD     &  &  \onlinecite{st92} & C$_{\rm s}$        & --   & EMT     \\             
  &  \onlinecite{do98}  & D$_{\rm 3h}$       & -- & SC        & &  \onlinecite{na97} & D$_{\rm 5h}$       & 2.28 & FS       &  &  \onlinecite{de98} & C$_{\rm s}$        & 2.28 & AI      \\             
  &  \onlinecite{lo94}  & D$_{\rm 3h}$       & -- & $n$G      & &  \onlinecite{bo01,bo_comm03}& D$_{\rm 5h}$ & 2.37 & EAM   &  &  \onlinecite{kr00} & C$_{\rm 2v}$       & 2.36 & AI      \\             
  &  \onlinecite{mi01}  & D$_{\rm 3h}$       & 2.26 & AI      & &  \onlinecite{st92} & D$_{\rm 5h}$       & --   & EMT      &  &  \onlinecite{pa94} & C$_{\rm s}$        & --   & exp     \\                
  &  \onlinecite{re95}  & C$_{\rm 2v}$       & 2.15 & AI      & &  \onlinecite{do98} & D$_{\rm 5h}$       & --   & SC       &  8.3& here                   & D$_{\rm 3d}$       & 2.39 & EAM \\          
  &  \onlinecite{pa94}  & D$_{\rm 3h}$       & --   & exp     & &  \onlinecite{lo94,mi99} & D$_{\rm 5h}$  & --   & $n$G     &  &  \onlinecite{bo_comm03}& D$_{\rm 3d}$      & 2.36 & EAM  \\             
4 & here                    & T$_{\rm d}$        & 2.32 & EAM & &  \onlinecite{de98} & C$_{\rm 3v}$       & 2.27 & AI       &  &  \onlinecite{de98} & C$_{\rm 2v}$       & 2.26 & AI      \\             
  &  \onlinecite{la96}  & D$_{\rm 4h}$       & 2.26 & TBMD    & &  \onlinecite{na96} & C$_{\rm 3v}$       & --   & AI       &  &  \onlinecite{kr00} & D$_{\rm 3d}$/C$_{\rm s}$  & 2.37/2.38 & AI \\  
  &  \onlinecite{na97}  & T$_{\rm d}$        & 2.20 & FS      & &  \onlinecite{pa94} & C$_{\rm 3v}$       & --   & exp      &  8.4& here                   & C$_{\rm 2v}$       & 2.39 & EAM \\          
  &  \onlinecite{st92}  & T$_{\rm d}$        & --   & EMT     & 7.2& here                   & C$_{\rm 3v}$& 2.38 & EAM      &  &  \onlinecite{bo_comm03}& C$_{\rm 2v}$   & 2.36 & EAM     \\             
  &  \onlinecite{do98}  & T$_{\rm d}$        & -- & SC        & &  \onlinecite{bo_comm03} & C$_{\rm 3v}$  & 2.36 & EAM      &  &  \onlinecite{de98} & D$_{\rm 2d}$       & 2.31 & AI      \\             
  &  \onlinecite{lo94}  & T$_{\rm d}$        & -- & $n$G      & &  \onlinecite{st92} & C$_{\rm 3v}$       & -- &  EMT       &  13.1& here                  & I$_{\rm h}$        & 2.36/2.48 & EAM \\ 
&\onlinecite{mi01}&D$_{\rm 2d}$ ($\sim$T$_{\rm d}$)&2.33 & AI & &  \onlinecite{mi99} & C$_{\rm 3v}$       & -- & $n$G       &  &  \onlinecite{la96} & I$_{\rm h}$       & 2.57 & TBMD    \\              
&\onlinecite{re95}&D$_{\rm 2d}$/D$_{\rm 4h}$& 2.17/2.10&AI& &\onlinecite{de98}&C$_{\rm 2v}$ ($\sim$D$_{\rm 5h}$) &2.28&AI  &  & \onlinecite{na97} & I$_{\rm h}$       & --/2.39 & FS \\              
  &  \onlinecite{pa94}  &    --              & --   & exp     & &  \onlinecite{na96} & D$_{\rm 5h}$       & --   & AI       &  &  \onlinecite{re96,bo01}& I$_{\rm h}$   & -- & EAM       \\              
5 & here                    & D$_{\rm 3h}$       & 2.35 & EAM & &  \onlinecite{pa94} & D$_{\rm 5h}$       & --   & exp      &  &  \onlinecite{lo94,mi99} & I$_{\rm h}$  & -- & $n$G      \\              
  &  \onlinecite{la96}  & T$_{\rm d} $       & 2.42 & TBMD    & 7.3& here                   & C$_{\rm 2}$      & 2.38 & EAM &  &  \onlinecite{st92} & I$_{\rm h}$        & --   & EMT    \\              
  &  \onlinecite{na97}  & D$_{\rm 3h}$       & 2.23 & FS      & &  \onlinecite{bo_comm03} & C$_{\rm 3v}$  & 2.36 & EAM      &  &  \onlinecite{do98} & I$_{\rm h}$        & --   & SC     \\                
  &  \onlinecite{st92}  & D$_{\rm 3h}$       & --   & EMT     & &  \onlinecite{st92} & C$_{\rm 2}$        & --   &  EMT     &  &  \onlinecite{re95} & I$_{\rm h}$        & 2.23/2.34 & AI \\           
  &  \onlinecite{do98}  & D$_{\rm 3h}$       & --   & SC      & &  \onlinecite{mi99} & C$_{\rm 2}$        & --   & $n$G     &  &  \onlinecite{ca99} & I$_{\rm h}$        & 2.41/2.53 & AI \\           
  &  \onlinecite{lo94}  & D$_{\rm 3h}$       & -- & $n$G      & &  \onlinecite{de98} & C$_{\rm 3v}$       & 2.28 & AI       &  &  \onlinecite{pa94} & I$_{\rm h}$        & --  & exp     \\              
  &  \onlinecite{mi01}  & C$_{\rm 4v}$       & 2.35 & AI      & 7.4& here                   & C$_{\rm 3v}$     & 2.39 & EAM &  13.2& here                   & C$_{\rm s}$        & 2.44 & EAM\\          
  &  \onlinecite{re95}  & D$_{\rm 3h}$       & 2.25 & AI      & &  \onlinecite{bo_comm03} & C$_{\rm 2}$   & 2.37 & EAM      &  &  \onlinecite{la96} & O$_{\rm h}$        & 2.48 & TBMD   \\              
  &  \onlinecite{pa94}  & D$_{\rm 3h}$       & --   & exp     & &  \onlinecite{de98} & C$_{\rm 2v}$       & 2.25 & AI       &  &  \onlinecite{st92} & D$_{\rm 3h}$       & --  & EMT     \\              
6.1& here                   & O$_{\rm h}$      & 2.36 & EAM   & 8.1&here                    & D$_{\rm 2d}$     & 2.38 & EAM &  &  \onlinecite{mi99} & C$_{\rm s}$        & --  & $n$G    \\              
  &  \onlinecite{la96} & D$_{\rm 4h}$       & 2.47 & TBMD     & &  \onlinecite{bo01,bo_comm03}& D$_{\rm 2d}$ & 2.36 & EAM   &  13.3& here                   & C$_{\rm s}$        & 2.44 & EAM\\          
  &  \onlinecite{na97} & O$_{\rm h}$        & 2.25 & FS       & &  \onlinecite{la96} & C$_{\rm 2h}$       & 2.50 & TBMD     &  &  \onlinecite{st92} & O$_{\rm h}$        & --  & EMT     \\              
  &  \onlinecite{st92} & O$_{\rm h}$        & --   & EMT      & &  \onlinecite{na97} & D$_{\rm 2d}$       & 2.25 & FS       &  &  \onlinecite{mi99} & C$_{\rm s}$        & --  & $n$G    \\ 
  &                    &                    &      &          & &  \onlinecite{st92} & D$_{\rm 2d}$       & --   & EMT      &  &  here              & C$_{\rm s}$        & 2.43 & EAM     \\              
\end{tabular}
\label{tab00}
\end{ruledtabular}
\end{table}
\endgroup

\begin{table}[tbp]
\begin{center}
\caption[kurzform]{Point groups of the optimized clusters.}
\begin{tabular}{cllllcllllcllllcllll}
\hline\hline
$\, N \, $  & $N.1$   &
  $N.2$  & $N.3$  & $N.4$ \quad &  $\, N \, $ &  $N.1$  &  $N.2$  & $N.3$ & $N.4$ \quad &  $\, N \, $ & $N.1$  &  $N.2$  & $N.3$ & $N.4$ &  $\, N \, $  & $N.1$  &  $N.2$  & $N.3$ & $N.4$  \\
\hline
2  & D$_{\rm \infty h}$&    &&                                 & 40 & C$_{\rm s}$ & C$_{\rm s}$ & C$_{\rm 1}$ & C$_{\rm 2}$ & 78 & C$_{\rm 1}$ & C$_{\rm 1}$ & C$_{\rm s}$ & C$_{\rm s}$ & 116& C$_{\rm 5v}$& C$_{\rm 1}$ & C$_{\rm 1}$ & C$_{\rm 1}$  \\ 
[-1.3ex]

3  & D$_{\rm 3h}$ &         &&                                 & 41 & C$_{\rm s}$ & C$_{\rm s}$ & C$_{\rm 1}$ & C$_{\rm 1}$ & 79 & C$_{\rm 2v}$& O$_{\rm h}$ & D$_{\rm 3h}$& C$_{\rm 1}$ & 117& C$_{\rm 1}$ & C$_{\rm s}$ & C$_{\rm 1}$ & C$_{\rm 1}$  \\ 
[-1.3ex]

4  & T$_{\rm d}$  &         &&                                 & 42 & C$_{\rm s}$ & C$_{\rm s}$ & C$_{\rm 1}$ & C$_{\rm s}$ & 80 & C$_{\rm 1}$ & C$_{\rm s}$ & C$_{\rm 2v}$& C$_{\rm s}$ & 118& C$_{\rm s}$ & C$_{\rm s}$ & C$_{\rm 1}$ & C$_{\rm 1}$   \\ [-1.3ex]

5  & D$_{\rm 3h}$ &         &&                                 & 43 & C$_{\rm s}$ & C$_{\rm s}$ & C$_{\rm 1}$ & C$_{\rm s}$ & 81 & C$_{\rm 2}$ & C$_{\rm s}$ & C$_{\rm s}$ & C$_{\rm 2v}$& 119& C$_{\rm s}$ & C$_{\rm 1}$ & C$_{\rm 1}$ & C$_{\rm 1}$   \\ [-1.3ex]

6  & O$_{\rm h}$  & C$_{\rm 2v}$  &&                           & 44 & C$_{\rm 1}$ & C$_{\rm s}$ & C$_{\rm 1}$ & C$_{\rm s}$ & 82 & C$_{\rm 1}$ & C$_{\rm s}$ & C$_{\rm s}$ & C$_{\rm 1}$ & 120& C$_{\rm 1}$ & C$_{\rm 1}$ & C$_{\rm 1}$ & C$_{\rm s}$   \\ [-1.3ex]

7  & D$_{\rm 5h}$ & C$_{\rm 3v}$ & C$_{\rm 2}$ & C$_{\rm 3v}$  & 45 & C$_{\rm s}$ & C$_{\rm s}$ & C$_{\rm 1}$ & C$_{\rm 1}$ & 83 & C$_{\rm 2v}$& D$_{\rm 3}$ & C$_{\rm 2}$ & C$_{\rm s}$ & 121& C$_{\rm 1}$ & C$_{\rm 1}$ & C$_{\rm 1}$ & C$_{\rm 1}$   \\ [-1.3ex]

8  & D$_{\rm 2d}$ & C$_{\rm s}$  & D$_{\rm 3d}$ & C$_{\rm 2v}$ & 46 & C$_{\rm 2v}$& C$_{\rm 1}$ & C$_{\rm 1}$ & C$_{\rm 1}$ & 84 & C$_{\rm s}$ & C$_{\rm 1}$ & C$_{\rm 1}$ & C$_{\rm 1}$ & 122& C$_{\rm 1}$ & C$_{\rm 1}$ & C$_{\rm s}$ & C$_{\rm 1}$  \\ [-1.3ex]

9  & C$_{\rm 2v}$ & D$_{\rm 3h}$ & C$_{\rm 2v}$ & C$_{\rm s}$  & 47 & C$_{\rm 1}$ & C$_{\rm s}$ & C$_{\rm s}$ & C$_{\rm s}$ & 85 & C$_{\rm 1}$ & C$_{\rm s}$ & C$_{\rm 1}$ & C$_{\rm 1}$ & 123& C$_{\rm s}$ & C$_{\rm s}$ & C$_{\rm 1}$ & C$_{\rm 1}$   \\ [-1.3ex]

10 & C$_{\rm 3v}$ & D$_{\rm 2h}$ & C$_{\rm 2}$  & C$_{\rm 2v}$ & 48 & C$_{\rm s}$ & C$_{\rm s}$ & C$_{\rm s}$ & C$_{\rm 1}$ & 86 & C$_{\rm 3}$ & C$_{\rm s}$ & C$_{\rm 1}$ & C$_{\rm 1}$ & 124& C$_{\rm s}$ & C$_{\rm s}$ & C$_{\rm s}$ & C$_{\rm 1}$   \\ [-1.3ex]

11 & C$_{\rm 2v}$ & C$_{\rm 2}$  & C$_{\rm 2v}$ & C$_{\rm 2}$  & 49 & C$_{\rm 3v}$& C$_{\rm 1}$ & C$_{\rm s}$ & C$_{\rm s}$ & 87 & C$_{\rm 1}$ & C$_{\rm 1}$ & C$_{\rm s}$ & C$_{\rm s}$ & 125& C$_{\rm s}$ & C$_{\rm 1}$ & C$_{\rm 1}$ & C$_{\rm 1}$    \\ [-1.3ex]

12 & C$_{\rm 5v}$ & C$_{\rm 1}$  & D$_{\rm 2d}$ & D$_{\rm 3h}$ & 50 & C$_{\rm s}$ & C$_{\rm s}$ & C$_{\rm 2v}$& C$_{\rm s}$ & 88 & C$_{\rm s}$ & C$_{\rm s}$ & C$_{\rm s}$ & C$_{\rm 1}$ & 126& C$_{\rm s}$ & C$_{\rm s}$ & C$_{\rm s}$ & C$_{\rm 1}$    \\ 
 [-1.3ex]

13 & I$_{\rm h}$  & C$_{\rm s}$  & C$_{\rm s}$  & C$_{\rm s}$  & 51 & C$_{\rm 2v}$& C$_{\rm s}$ & C$_{\rm s}$ & C$_{\rm 2}$ & 89 & C$_{\rm 3v}$& C$_{\rm s}$ & C$_{\rm 1}$ & C$_{\rm s}$ & 127& C$_{\rm 2v}$& C$_{\rm s}$ & C$_{\rm 1}$ & C$_{\rm 1}$    \\  [-1.3ex]

14 & C$_{\rm 3v}$ & C$_{\rm 2v}$ & C$_{\rm s}$  & C$_{\rm 2v}$ & 52 & C$_{\rm 3v}$& C$_{\rm s}$ & C$_{\rm s}$ & C$_{\rm s}$ & 90 & C$_{\rm s}$ & C$_{\rm 1}$ & C$_{\rm s}$ & C$_{\rm 1}$ & 128& C$_{\rm 1}$ & C$_{\rm 1}$ & C$_{\rm s}$ & C$_{\rm s}$    \\  [-1.3ex]

15 & C$_{\rm 2v}$ & D$_{\rm 6d}$ & C$_{\rm 2v}$ & C$_{\rm 2v}$ & 53 & C$_{\rm 2v}$& D$_{\rm 5d}$& C$_{\rm 2v}$& C$_{\rm s}$ & 91 & C$_{\rm s}$ & C$_{\rm s}$ & C$_{\rm s}$ & C$_{\rm 1}$ & 129& C$_{\rm s}$ & C$_{\rm s}$ & C$_{\rm s}$ & C$_{\rm 2}$    \\  [-1.3ex]

16 & C$_{\rm s}$  & C$_{\rm s}$  & C$_{\rm 2}$  & D$_{\rm 3h}$ & 54 & C$_{\rm 5v}$& I$_{\rm h}$ & C$_{\rm 2v}$& C$_{\rm s}$ & 92 & C$_{\rm 3v}$&      T      & C$_{\rm 1}$ & C$_{\rm 1}$ & 130& C$_{\rm 1}$ & C$_{\rm s}$ & C$_{\rm s}$ & C$_{\rm 1}$    \\ 
 [-1.3ex]

17 & C$_{\rm 2}$  & C$_{\rm s}$  & C$_{\rm s}$  & C$_{\rm 3v}$ & 55 & I$_{\rm h}$ & C$_{\rm s}$ & C$_{\rm s}$ & C$_{\rm s}$ & 93 & C$_{\rm 1}$ & C$_{\rm 1}$ & C$_{\rm 1}$ & C$_{\rm 1}$ & 131& C$_{\rm 2v}$& C$_{\rm 1}$ & C$_{\rm 1}$ & C$_{\rm 1}$    \\ 
 [-1.3ex]

18 & C$_{\rm s}$  & C$_{\rm 5v}$ & C$_{\rm s}$  & C$_{\rm 1}$  & 56 & C$_{\rm 3v}$& C$_{\rm 3v}$& C$_{\rm s}$ & C$_{\rm 1}$ & 94 & C$_{\rm 1}$ & C$_{\rm 1}$ & C$_{\rm 1}$ & C$_{\rm 1}$ & 132& C$_{\rm 1}$ & C$_{\rm 1}$ & C$_{\rm s}$ & C$_{\rm 1}$    \\ 
 [-1.3ex]

19 & D$_{\rm 5h}$ & C$_{\rm 1}$  & C$_{\rm s}$  & C$_{\rm 1}$  & 57 & C$_{\rm s}$ & C$_{\rm s}$ & C$_{\rm s}$ & C$_{\rm s}$ & 95 & C$_{\rm 1}$ & C$_{\rm 3}$ & C$_{\rm 1}$ & C$_{\rm 1}$ & 133& C$_{\rm 1}$ & C$_{\rm s}$ & C$_{\rm s}$ & C$_{\rm 1}$    \\ 
 [-1.3ex]

20 & C$_{\rm 2v}$ & D$_{\rm 3d}$ & D$_{\rm 2}$  & D$_{\rm 2}$  & 58 & C$_{\rm 3v}$& C$_{\rm s}$ & C$_{\rm 1}$ & C$_{\rm 1}$ & 96 & C$_{\rm 1}$ & C$_{\rm 1}$ & C$_{\rm 1}$ & C$_{\rm 2v}$& 134& C$_{\rm 3v}$& C$_{\rm s}$ & C$_{\rm 1}$ & C$_{\rm 1}$    \\ 
 [-1.3ex]

21 & C$_{\rm 1}$  & C$_{\rm 2v}$ & C$_{\rm 2v}$ & C$_{\rm s}$  & 59 & C$_{\rm 2v}$& C$_{\rm 1}$ & C$_{\rm 1}$ & C$_{\rm 2v}$& 97 & C$_{\rm 1}$ & C$_{\rm 1}$ & C$_{\rm 1}$ & C$_{\rm 2v}$& 135& C$_{\rm s}$ & C$_{\rm 1}$ & C$_{\rm s}$ & C$_{\rm s}$    \\  [-1.3ex]

22 & C$_{\rm s}$  & C$_{\rm s}$  & C$_{\rm 1}$  & C$_{\rm s}$  & 60 & C$_{\rm s}$ & C$_{\rm s}$ & C$_{\rm s}$ & C$_{\rm 1}$ & 98 & C$_{\rm s}$ & C$_{\rm s}$ & C$_{\rm s}$ & C$_{\rm s}$ & 136& C$_{\rm s}$ & C$_{\rm s}$ & C$_{\rm s}$ & C$_{\rm 1}$    \\  [-1.3ex]

23 & D$_{\rm 3h}$ & C$_{\rm 2}$  & C$_{\rm s}$  & C$_{\rm 1}$  & 61 & C$_{\rm 2v}$& C$_{\rm s}$ & C$_{\rm s}$ & C$_{\rm 1}$ & 99 & C$_{\rm 2v}$& C$_{\rm 1}$ & C$_{\rm s}$ & C$_{\rm s}$ & 137& C$_{\rm 3v}$& C$_{\rm 2v}$& C$_{\rm 2v}$& C$_{\rm 1}$     \\  [-1.3ex]

24 & C$_{\rm 2v}$ & C$_{\rm s}$  & D$_{\rm 3}$  & C$_{\rm s}$  & 62 & C$_{\rm 2v}$& C$_{\rm s}$ & C$_{\rm 1}$ & C$_{\rm 1}$ & 100& C$_{\rm s}$ & C$_{\rm s}$ & C$_{\rm 5v}$& C$_{\rm 1}$ & 138& C$_{\rm 3v}$& C$_{\rm s}$ & C$_{\rm 1}$ & C$_{\rm s}$     \\ 
 [-1.3ex]

25 & C$_{\rm 2v}$ & C$_{\rm 3}$  & D$_{\rm 5d}$ & C$_{\rm s}$  & 63 & C$_{\rm 1}$ & C$_{\rm s}$ & C$_{\rm 1}$ & C$_{\rm s}$ & 101& C$_{\rm 2v}$& D$_{\rm 5h}$& C$_{\rm s}$ & C$_{\rm 1}$ & 139& C$_{\rm 2v}$& C$_{\rm s}$ & C$_{\rm s}$ & C$_{\rm s}$     \\ 
 [-1.3ex]

26 & T$_{\rm d}$  & C$_{\rm s}$  & C$_{\rm 1}$  & C$_{\rm s}$  & 64 & C$_{\rm s}$ & C$_{\rm 1}$ & C$_{\rm 1}$ & C$_{\rm 2v}$& 102& C$_{\rm s}$ & C$_{\rm 1}$ & C$_{\rm 1}$ & C$_{\rm 2v}$& 140& C$_{\rm s}$ & C$_{\rm s}$ & C$_{\rm 1}$ & C$_{\rm s}$     \\ 
 [-1.3ex]

27 & C$_{\rm 2v}$ & C$_{\rm s}$  & C$_{\rm s}$  & C$_{\rm s}$  & 65 & C$_{\rm 1}$ & C$_{\rm 2}$ & C$_{\rm 1}$ & C$_{\rm 1}$ & 103& C$_{\rm 1}$ & C$_{\rm s}$ & C$_{\rm 1}$ & C$_{\rm 1}$ & 141& C$_{\rm 5v}$& C$_{\rm 3v}$& C$_{\rm 2}$ & C$_{\rm 1}$    \\ 
 [-1.3ex]

28 & T            & D$_{\rm 2}$  & C$_{\rm 3v}$ & C$_{\rm s}$  & 66 & C$_{\rm 1}$ & C$_{\rm 1}$ & C$_{\rm 1}$ & C$_{\rm 1}$ & 104& C$_{\rm 2v}$& C$_{\rm 1}$ & C$_{\rm s}$ & C$_{\rm 1}$ & 142& C$_{\rm s}$ & C$_{\rm s}$ & C$_{\rm 5v}$& C$_{\rm s}$    \\ 
 [-1.3ex]

29 & C$_{\rm 3}$  & D$_{\rm 3h}$ & C$_{\rm 2v}$ & C$_{\rm s}$  & 67 & C$_{\rm 2}$ & C$_{\rm s}$ & C$_{\rm 2v}$& C$_{\rm s}$ & 105& C$_{\rm 1}$ & C$_{\rm 1}$ & C$_{\rm 1}$ & C$_{\rm s}$ & 143& C$_{\rm 2v}$& C$_{\rm s}$ & C$_{\rm s}$ & C$_{\rm s}$   \\ 
 [-1.3ex]

30 & C$_{\rm 2v}$ & C$_{\rm s}$  & C$_{\rm 1}$  & C$_{\rm 1}$  & 68 & C$_{\rm 1}$ & C$_{\rm 1}$ & C$_{\rm 1}$ & C$_{\rm 2}$ & 106& C$_{\rm 1}$ & C$_{\rm 1}$ & C$_{\rm 1}$ & C$_{\rm 1}$ & 144& C$_{\rm 3v}$& C$_{\rm s}$ & C$_{\rm s}$ & C$_{\rm s}$   \\ 
 [-1.3ex]

31 & C$_{\rm s}$  & C$_{\rm 3}$  & C$_{\rm s}$  & C$_{\rm 1}$  & 69 & C$_{\rm 1}$ & C$_{\rm 1}$ & C$_{\rm 5v}$& C$_{\rm 1}$ & 107& C$_{\rm s}$ & C$_{\rm 1}$ & C$_{\rm 1}$ & C$_{\rm s}$ & 145& C$_{\rm 2v}$& C$_{\rm 2v}$& D$_{\rm 5d}$& C$_{\rm s}$  \\ 
 [-1.3ex]

32 & D$_{\rm 3}$  & C$_{\rm 2v}$ & C$_{\rm 2v}$ & C$_{\rm s}$  & 70 & C$_{\rm 5v}$& C$_{\rm 1}$ & C$_{\rm 5}$ & C$_{\rm 5}$ & 108& C$_{\rm s}$ & C$_{\rm 1}$ & C$_{\rm 1}$ & C$_{\rm 1}$ & 146& C$_{\rm 5v}$& C$_{\rm s}$ & C$_{\rm s}$ & C$_{\rm s}$  \\ 
 [-1.3ex]

33 & C$_{\rm 2}$  & C$_{\rm s}$  & C$_{\rm s}$  & C$_{\rm s}$  & 71 & C$_{\rm 5}$ & C$_{\rm 5v}$& C$_{\rm 2v}$& C$_{\rm s}$ & 109& C$_{\rm 1}$ & C$_{\rm s}$ & C$_{\rm 1}$ & C$_{\rm 1}$ & 147& I$_{\rm h}$ & C$_{\rm 1}$ & C$_{\rm 1}$ & C$_{\rm 1}$   \\ 
 [-1.3ex]

34 & C$_{\rm s}$  & C$_{\rm s}$  & C$_{\rm s}$  & C$_{\rm 1}$  & 72 & C$_{\rm s}$ & C$_{\rm s}$ & C$_{\rm s}$ & C$_{\rm 1}$ & 110& C$_{\rm s}$ & C$_{\rm s}$ & C$_{\rm 1}$ & C$_{\rm 1}$ & 148& C$_{\rm s}$ & C$_{\rm s}$ & C$_{\rm s}$ & C$_{\rm 1}$   \\ 
 [-1.3ex]

35& C$_{\rm 2v}$  & D$_{\rm 3}$  & D$_{\rm 3}$  & C$_{\rm s}$  & 73 & C$_{\rm s}$ & C$_{\rm s}$ & C$_{\rm s}$ & C$_{\rm 1}$ & 111& C$_{\rm s}$ & C$_{\rm s}$ & C$_{\rm 1}$ & C$_{\rm 1}$ & 149& C$_{\rm s}$ & C$_{\rm s}$ & C$_{\rm s}$ & C$_{\rm 1}$   \\ 
 [-1.3ex]

36& C$_{\rm s}$   & C$_{\rm 1}$  & C$_{\rm 1}$  & C$_{\rm 1}$  & 74 & C$_{\rm 5v}$& C$_{\rm s}$ & C$_{\rm 1}$ & C$_{\rm 1}$ & 112& C$_{\rm s}$ & C$_{\rm 1}$ & C$_{\rm s}$ & C$_{\rm 1}$ & 150& C$_{\rm 3v}$& C$_{\rm 3v}$ & C$_{\rm s}$ & C$_{\rm 1}$   \\ 
 [-1.3ex]

37& C$_{\rm 2}$   & C$_{\rm 1}$  & C$_{\rm 1}$  & C$_{\rm 1}$  & 75 & D$_{\rm 5h}$& C$_{\rm s}$ & C$_{\rm s}$ & C$_{\rm s}$ & 113& C$_{\rm s}$ & C$_{\rm s}$ & C$_{\rm 1}$ & C$_{\rm 1}$   \\ 
 [-1.3ex]

38& O$_{\rm h}$   & C$_{\rm 5v}$ & C$_{\rm 5v}$ & C$_{\rm 1}$  & 76 & C$_{\rm s}$ & C$_{\rm 2v}$& C$_{\rm s}$ & C$_{\rm 1}$ & 114& C$_{\rm s}$ & C$_{\rm s}$ & C$_{\rm 5v}$& C$_{\rm 1}$  \\ 
 [-1.3ex]

39& C$_{\rm 5v}$  & C$_{\rm 5}$  & C$_{\rm s}$  & C$_{\rm 4v}$ & 77 & C$_{\rm 2v}$& C$_{\rm 2}$ & C$_{\rm s}$ & C$_{\rm 2}$ & 115& C$_{\rm 5v}$& C$_{\rm s}$ & C$_{\rm 5v}$& C$_{\rm 1}$  \\

\hline\hline
\end{tabular}
\label{I}
\end{center}
\end{table}

\unitlength1cm

\unitlength1cm

\unitlength1cm

\unitlength1cm

\unitlength1cm

\unitlength1cm

\unitlength1cm

\unitlength1cm

\unitlength1cm

\unitlength1cm

\unitlength1cm


\begin{thebibliography}{99} 

\bibitem{tsai93} 
C. J. Tsai and K. D. Jordan,  J. Phys. Chem. {\bf 97}, 11227 (1993).

\bibitem{ca99} 
M.  Calleja, C.  Rey, M. M. G.  Alemany, L. J.  Gallego, P.
Ordej\'on, D.  S\'anchez-Portal, E.  Artacho and J. M.  Soler,  Phys.
Rev. B {\bf 60}, 2020 (
1999).

\bibitem{mi01} 
M. C. Michelini, R. Pis Diez, and A. H. Jubert,  Int. J. Quant. Chem.
{\bf 85}, 22 (2001).

\bibitem{re95}
F. A. Reuse and S. N. Khanna,
Chem. Phys. Lett. {\bf 234}, 77 (1995).

\bibitem{na96}
S. K. Nayak, B. Reddy, B. K. Rao, S. N. Khanna, and P. Jena,
Chem. Phys. Lett. {\bf 253}, 390 (1996).

\bibitem{de98} N. Desmarais, C. Jamorski, F. A. Reuse, and S. N. Khanna,
 Chem. Phys. Lett. {\bf 294}, 480 (1998).

\bibitem{kr00} S. Kr\"uger, T. J. Seem\"uller, A. W\"orndle, and 
N. R\"osch,  Int. J. Quant. Chem. {\bf 80}, 567 (2000).

\bibitem{vl92}
D. G. Vlachos, L. D. Schmidt, and R. Aris,  J. Chem. Phys. {\bf 96}, 
6880 (1992).

\bibitem{re93}
C. Rey, L. J. Gallego, J. Garc\'{\i}a-Rodeja, J. A. Alonso and 
M. P. I\~{n}iguez,  Phys. Rev. B {\bf 48}, 8253 (1993).

\bibitem{mo94}
J. M. Montejano-Carrizales, M. P. I\~{n}iguez, and J. A. Alonso,
 J. Cluster. Sci. {\bf 5}, 287  (1994).

\bibitem{mo96}
J. M. Montejano-Carrizales, M. P. I\~{n}iguez, J. A. Alonso, and M. J. L\'{o}pez,
 Phys. Rev. B {\bf 54}, 5961  (1996).

\bibitem{ga00}
L. Garc\'{\i}a Gonz\'{a}lez and J. M. Montejano-Carrizales,
 phys. stat. sol. {\bf 220}, 357 (2000).

\bibitem{gu92} 
Z. B. G\"uven\c{c}, J. Jellinek and A. F. Voter, in {\it Physics and Chemistry 
of Finite systems: From Clusters to Crystals}, edited by P. Jena, S. N. Khanna
and B. K. Rao (Kluwer, Dordrecht, 1992), Vol I, p. 411.

\bibitem{je93}
Z. B. G\"uven\c{c} and J. Jellinek,  Z. Phys. D {\bf 26}, 304 (1993).

\bibitem{bo01}
M. B\"oy\"ukata, Z. B. G\"uven\c{c}, S. \"Oz\c{c}elik, P. Durmus, and J. Jellinek,
 Int. J. Quant. Chem. {\bf 84}, 208 (2001).

\bibitem{st92}
M. S. Stave and A. E. DePristo,  J. Chem. Phys. {\bf 97}, 3386  (1992).

\bibitem{we96}
T. L. Wetzel and A. E. DePristo,  J. Chem. Phys. {\bf 105}, 572 (1996).

\bibitem{je91}
J. Jellinek and I. L. Garz\'{o}n,  Z. Phys. D {\bf 20}, 239 (1991).

\bibitem{ga92}
I. L. Garz\'{o}n and  J. Jellinek, in {\it Physics and Chemistry 
of Finite systems: From Clusters to Crystals}, ed. P. Jena, S.N. Khanna
and B.K. Rao, Kluwer, Dordrecht, 1992, Vol I, p. 405.

\bibitem{lo94}
M. J. Lopez and J. Jellinek,  Phys. Rev. A {\bf 50}, 1445 (1994).

\bibitem{ag98}
F. Aguilera-Granja, S. Bouarab, M. J. L\'opez, A. Vega, 
J. M. Montejano-Carrizales, M. P. I\~{n}iguez, and J. A. Alonso,
 Phys. Rev. B {\bf 57}, 12469 (1998).

\bibitem{mi99} 
K. Michaelian, N. Rend\'on, and I. L. Garz\'on,
 Phys. Rev. B {\bf 60}, 2000 (1999).

\bibitem{do98}
J. P. K. Doye and D. J. Wales, New J. Chem. {\bf 22}, 733 (1998).

\bibitem{northby}
J. A. Northby, J. Chem. Phys. {\bf 87}, 6166 (1987).

\bibitem{na97} 
S. K. Nayak, S. N. Khanna, B. K. Rao, and P. Jena, 
 J. Chem. Phys. A {\bf 101}, 1072 (1997).

\bibitem{ll00} 
L. D. Lloyd, R. L. Johnston, 
 J. Chem. Soc., Dalton Trans. {\bf 3}, 307 (2000).

\bibitem{hu96} 
W. Hu, L. M. Mei, and H. Li,
 Solid State Commun. {\bf 100}, 129 (1996).

\bibitem{la96} 
N. N. Lathiotakis, A. N. Andriotis, M. Menon and J. Connolly,
 J. Chem. Phys. {\bf 104}, 992  (1996).

\bibitem{pa91}
E. K. Parks, B. J. Winter, T. D. Klots and S. J. Riley, 
 J. Chem. Phys. {\bf 94}, 1882 (1991).

\bibitem{pa94}
E. K. Parks, L. Zhu, J. Ho and S. J. Riley,
 J. Chem. Phys. {\bf 100}, 7206 (1994).

\bibitem{pa95}
E. K. Parks, L. Zhu, J. Ho and S. J. Riley, 
 J. Chem. Phys. {\bf 102}, 7377 (1995).

\bibitem{pa95a}
E. K. Parks and S. J. Riley 
 Z. Phys. D. {\bf 33},  59, (1995).

\bibitem{pa97}
E. K. Parks, G. C. Nieman, K. P. Kerns, and S. J. Riley, 
 J. Chem. Phys.  {\bf 107}, 1861 (1997).

\bibitem{pa98}
E. K. Parks, K. P. Kerns, and S. J. Riley, 
 J. Chem. Phys.  {\bf 109}, 10207 (1998).

\bibitem{pa01}
E. K. Parks, K. P. Kerns, and S. J. Riley, 
 J. Chem. Phys.  {\bf 114}, 2228 (2001).

\bibitem{kn01}
M. B. Knickelbein, 
 J. Chem. Phys.  {\bf 115}, 5957 (2001).

\bibitem{pe94}
M. Pellarin, B. Baguenard, J. L. Vialle, J. Lerme, M. Broyer,
J. Miller and A. Perez,
 Chem. Phys. Lett. {\bf 217}, 349 (1994).

\bibitem{pccp01} 
V. G. Grigoryan and M. Springborg, Phys. Chem. Chem. Phys.
{\bf 3}, 5125 (2001).

\bibitem{cpl03} 
V. G. Grigoryan and M. Springborg, Chem. Phys. Lett.
{\bf 375}, 219 (2003).

\bibitem{da83} 
M. S. Daw and M. I. Baskes,  Phys.  Rev.  Lett. {\bf 50}, 1285 (1983).

\bibitem{da84} 
M. S. Daw and M. I. Baskes,  Phys.  Rev. B {\bf 29}, 6443 (
1984).

\bibitem{fo86} 
S. M. Foiles, M. S. Daw and M. I. Baskes,  Phys. Rev. B {\bf 33},
 7983 (1986).

\bibitem{da93}
M. S. Daw, S. M. Foiles and M. I. Baskes,  Mat. Sci. Rep. {\bf 9}, 251 
(1993).

\bibitem{cl74}
E. Clementi and C. Roetti,  At. Data Nucl. Data Tables {\bf 14}, 177
(1974).

\bibitem{mc81}
A.D. McLean and R.S. McLean, At. Data Nucl. Data Tables {\bf 26}, 197 (1981).

\bibitem{ro84} 
J.H. Rose, J.R. Smith, F. Guinea and J. Ferrante,  Phys.  Rev. B
{\bf 29}, 2963 (1984).

\bibitem{internet}
Internet address: 146.246.250.1

\bibitem{numer92}
W. H. Press, S. A. Teukolsky, W. T. Vetterling, B. P. Flannery, in
{\it Numerical Recipes in FORTRAN: the Art of Scientific Computing}, 
(Cambridge University Press, Cambridge, 1992), p. 387.

\bibitem{wa97} 
D. J. Wales and J. P. K. Doye,  J. Phys. Chem. A {\bf 101}, 5111 (1997).

\bibitem{pi95}
J. C. Pinegar, J. D. Langenberg, C. A. Arrington, E. M. Spain, and M. D. Morse,
J. Chem. Phys. {\bf 102}, 666 (1995).

\bibitem{bo_comm03}
Z. B. G\"uven\c{c} and M. B\"oy\"ukata (private communication).

\bibitem{cu98}
E. Curotto, A. Matro, D. L. Freeman, and J. D. Doll,
J. Chem. Phys. {\bf 108}, 729 (1998).

\bibitem{re96}
C. Rey, J. Garc\'{\i}a-Rodeja and L.J. Gallego,
 Phys. Rev. B {\bf 54}, 2942 (1996).

\bibitem{unpub}
V. G. Grigoryan and M. Springborg, unpublished results that can be obtained upon request
from the authors.

\bibitem{al00}
J. A. Alonso,
 Chem. Rev. {\bf 100}, 637 (2000).

\bibitem{do95}
J. P. K. Doye, D. J. Wales, and R. S. Berry,   J. Chem. Phys. {\bf 103}, 
4234 (1995).

\bibitem{ha84}
I. A. Harris, R. S. Kidwell, and J. A. Northby, 
Phys. Rev. Lett. {\bf 53},
 2390 (1984).

\bibitem{su94}
M. W. Sung, R. Kawai, and J. H. Weare, 
Phys. Rev. Lett. {\bf 73},
 3552 (1994).

\bibitem{cl91}
C. L. Cleveland and U. Landman,  J. Chem. Phys. {\bf 94}, 7376 (1991).

\bibitem{ma91}
T. P. Martin, T. Bergmann, H. G\"olich, and T. Lange,
Chem. Phys. Lett. {\bf 176}, 343 (1991).

\bibitem{vo87}
A. F. Voter and S. P. Chen, in {\it Characterization of Defects in Materials},
edited by R. W. Siegal, J. R. Weertman, and R. Sinclair, MRS Symposia 
Proceedings No. 82 (Materials Research Society, Pittsburgh, 1987), p. 175. 

\bibitem{vo95}
A. F. Voter, in {\it Intermetallic Compounds}, edited by J. H. Westbrook and 
R. L. Fleischer (John Wiley and Sons, Ltd, 1995), Vol. 1, p. 77.

\end{thebibliography}
\end{document}